# Motivated Reasoning and Blame: Responses to Performance Framing and Outgroup Triggers during COVID-19

Gregory Porumbescu,
*School of Public Affairs and Administration, Rutgers University*

Donald Moynihan,
*McCourt of Public Policy, Georgetown University*

Jason Anastasopoulos,
*School of Public and International Affairs, University of Georgia*

Asmus Leth Olsen,
*Department of Political Science, University of Copenhagen*

## Abstract

To manage citizen evaluations of government performance, public officials use blame avoidance strategies when communicating performance information. We examine two prominent presentational strategies: scapegoating and spinning, while testing how public responses vary depending on whether they are ideologically aligned with the public official. We examine these relationships in the context of the COVID-19 pandemic, where the Trump administration sought to shift blame by scapegoating outgroups (by using the term "Chinese virus"), and framing performance information on COVID-19 testing in positive terms. Using a novel pre-registered survey experiment that incorporates open and close-ended items, we offer three main findings. First, there is clear evidence of motivated reasoning: conservatives rate the performance of the Trump administration more positively and are more apt to blame prominent Democrats, Chinese residents and the Chinese Government. Second, performance information framing was found to impact blame attribution among conservatives, but only for open-ended responses. Third, while exposure to the term "Chinese virus" increased blame assigned to Chinese residents among all participants, conservatives exposed to the term appeared to blame President Trump more, suggesting repeated use of divisive blame shifting strategies may alienate even supporters.

A close-up photograph of Trump at the podium in mid-March made clear that he had used a Sharpie pen to replace the term "Coronavirus" with "Chinese virus" (Karni 2020). Across subsequent press conferences, rallies and tweets, Trump and White House officials repeatedly invoked the term "Chinese virus," and less frequently, variants such as "Wuhan virus," or "Kung Flu." It wasn't just the Trump administration; the Republican Party sent memos to its candidates urging them to blame China (Insenstadt 2020). Around the same time, Asian-Americans reported increased racial hostility (Kandil 2020).

These actions by the Trump administration can be read as a classic if extraordinary attempt to shift blame away from what was broadly seen as a poor administrative response to the pandemic. A pervasive negativity bias in the electorate (Lau 1982; Soroka 2014) implies that elected officials have strong incentives to redirect blame through the use of presentational strategies that seek to manage perceptions of government performance (Hood 2010; Weaver 1986). Information asymmetries between the government and public coupled with the complexity of public service provision facilitate efforts to avoid blame by making it difficult for the public to identify what went wrong, and who was responsible (Lupia 1994; Swindell and Kelly 2002; James, Jilke, Petersen and Van de Walle 2016).

In this paper, we examine two presentational blame avoidance strategies that governments employ to manage perceptions of performance. The first is to "spin" information about an event to avoid blame (Hood 2010). The second is to deflect blame through the use of a scapegoat (Weaver 1986). While scapegoats in blame avoidance studies are typically other policy actors – e.g. ministers, opposition, bureaucrats, or private service providers – we extend research on blame shifting presentational strategies by focusing on ethnic outgroups evoked by the term "Chinese virus."

Episodes of blame avoidance are fundamentally debates about government performance, making performance information central to such discussions (Nielsen and Baekgard 2015; Nielsen and Moynihan 2017; Petersen, Laumann and Jakobsen 2019). In addition to using the term "Chinese virus," another Trump administration strategy was to frame how performance was discussed – consistent with Hood's (2010) characterization of spinning as a presentational strategy. When Trump repeatedly (and falsely) insisted that anyone who wanted a virus test could have one, he was directing attention to a very specific metric of government performance – testing availability, rather than death rates, hospitalizations, or the reproduction of the virus – and framing it in positive terms. Do these presentational strategies change how the public assigned blame for government performance? We address this question in the context of motivated reasoning.

The COVID-19 pandemic resulted in mass death and a fundamental test of governance capacity and leadership skill, offering a perfect setting to study the effects of presentational strategies as a tool for blame avoidance. In the United States, a deeply polarized political environment appeared to undermine efforts toward a coordinated civic and government response (Allcott et al. 2020; Kettl 2020). Daily press conferences convened by elected officials and ubiquitous media coverage regularly exposed the public to large amounts of performance information and illustrated the presentational blame avoidance strategies described above. COVID-19 is hereby a key case where presentational strategies for blame avoidance are routinely used on a daily basis by politicians and public managers.

This study uses the case of the COVID-19 pandemic to advance our understanding of how three key theoretical variables shape the impact of presentational strategies on the public's willingness to assign blame and who they assign blame to, when asked to evaluate government performance. The first is motivated reasoning, where we generally expect conservatives to be more sympathetic to a Republican President, and more responsive to his efforts to avoid blame. The second is the framing of performance information, where we examine if positive framing of a key performance metric, the availability of COVID-19 tests, reduces blame. Third, we examine the effect of using a scapegoat cue – the use of the term Chinese virus – when conveying performance information. This attempt to scapegoat relies on people's inherent tendency to credit groups they identify with for positive events and blame groups they do not identify with for negative events (Pettigrew 1979; Gilbert and Malone 1995; Swim and Sanna 1996).

Our empirical analysis offers important theoretical contributions to public management literature on blame avoidance. Previous work has considered how motivated reasoning affects how the public interprets crises (Bisgaard 2015) and how they process performance information on health policy issues (James and Van Ryzin 2017). Our contribution comes from offering insight into how the framing of performance information and use of scapegoat cues in a highly polarized context shapes the way the public processes government performance information to attribute responsibility. Existing studies have not considered the role of ethnic-group characteristics in shaping patterns of performance evaluation and responsibility attribution. Such considerations are especially salient for tasks where outcomes are co-produced by the public and

government – such as halting a contagious disease – because cooperation appears less likely to emerge if some groups are scapegoated.

To make these contributions we use a novel pre-registered research design that leverages a survey experiment of US residents[1]. Additionally, we employ machine learning techniques to analyze how open-ended responses vary in relation to the performance information participants were randomly assigned to. This approach helps to elaborate upon the close-ended responses (Roberts et al 2014), while responding to recent calls for public administration research to incorporate machine learning (Anastasopoulos and Whitford 2019).

Three key findings emerge from our analysis. First, we find evidence of motivated reasoning. Conservatives were less likely to see the COVID-19 crisis as a serious threat, a performance failure, or to attribute blame to a Republican President and administration. Instead, they were more likely to blame targets of the President's criticism: former President Obama, Speaker Nancy Pelosi, the media, state governments, the Chinese government and Chinese-Americans. These findings are supported in an analysis of the open-ended responses.

Second, we show that using a scapegoat cue as a presentational strategy generally induces outgroup blaming. Participants exposed to the term Chinese virus were more apt to blame ethnic Chinese residents for the spread of COVID-19 in the United States relative to other ethnic groups. Notably, there is no effect of the scapegoat cue on blame assigned to the Chinese government. We also find that conservatives exposed to the term

Chinese virus were more likely to assign blame to public officials generally, including President Trump, creating a potential backlash for those using the term.

Third, we find evidence of a significant impact of the performance information framing presentational strategy for open-ended, but not close-ended items. These results help to shed light on the way individuals reason using performance information, suggesting that in complex, rapidly evolving contexts, individuals are susceptible to performance information framing, yet also reflect on the data before assigning blame. Specifically, while extant work on performance information framing documents a negativity bias in responses to performance information (Nielsen and Moynihan, 2017; Van Bekerom, van der Voet, and Christensen 2020), we show that blame attribution is not only impacted by negatively-, but also positively-framed performance information, and that who actors blame depends on the frame they are exposed to.

Next, we establish our expectations about how the key variables we examine shape responsibility attribution for public sector performance before providing more detail about the data and analysis.

*Performance Information and Responsibility Attribution*

Responsibility attribution results from speculation on the causes of performance an individual has observed (Heider 1958; Jones and Davis 1965). That is, exposure to performance information not only triggers performance evaluations, but also attributions of responsibility, especially for poor performance (Olsen 2015). However, the link between performance information and responsibility attribution is contested (Redlawsk 2002). While many theories assume individuals will use evidence to objectively and accurately assign blame for poor performance (Gerber and Green 1998), information users often fall short of this standard.

How the performance of public officials is evaluated, and how their presentational blame avoidance strategies are received is influenced by their perceived ideological alignment with their audience. Partisans engage in motivated reasoning to find and interpret information that limits blame to co-partisans and shifts blame to actors, such as politicians, they disagree with (Lodge and Taber 2000). In other words, rather than using performance information to arrive at more accurate conclusions, individuals are motivated to reason in a direction that confirms predetermined conclusions and aligns with preexisting biases (Olsen, Moynihan, James and Van Ryzin 2020).

The processing of performance information for attribution might also be driven by cognitive biases. These biases can be accentuated by presentational strategies. For example, negativity bias can be triggered by presenting numbers in negative, rather than positive, terms (Olsen 2015). Biases against outgroups might be activated by racial cues that trigger racial resentment (Whitehead, Smith, and Eichhorn 1982). The potential for scapegoating members of outgroups seems especially relevant for public sector tasks that are co-produced, i.e., where government relies on efforts from the public to achieve its goals. If accountability is driven by attribution processes to those perceived as responsible for outcomes, co-produced outcomes involve not just vertical models of accountability – blaming public officials – but also horizontal models – blaming other members of the public. In sum, the COVID-19 setting is ripe for scapegoat triggers to shape blame attribution for performance outcomes.

We examine each of these processes in turn.

*Partisan Motivated Reasoning and Blame Attribution*

Motivated reasoning suggests citizens evaluate members of the political party they identify with more positively and are more critical of the performance of parties they oppose (Taber and Lodge 2006). Jilke and Baekgaard (2020) show that citizen satisfaction with public services increases or decreases depending on whether co-partisans are in power. While crises can sometimes engender a tendency to "rally around the flag," political ideology offers a heuristic by which individuals make sense of crises where the situation is dynamic and the facts are contested (Bisgaard 2015). For example, liberal voters tended to blame a Republican President after the poor response to Hurricane Katrina, while Republicans blamed a Democratic governor (Malhotra and Kuo 2008).

Given that the claims of motivated reasoning are relatively well-established and not theoretically novel, our goal here is to simply observe if such processes hold in the COVID-19 case. More specifically, we expect conservatives to differ from the rest of the population significantly in how they interpret the crisis, with conservatives less likely to acknowledge the scale of the threat and more likely to positively evaluate the performance of Republican political leadership. We also expect conservatives to be more likely to attribute responsibility toward actors that President Trump has persistently sought to shift blame towards, such as President Obama, the mainstream media, state Governors, the Chinese government, and Chinese residents of the United States.

*H1: Motivated reasoning will lead to differences in how conservatives and non-conservatives assess the severity of the pandemic, assess the performance of political leadership, and allocate responsibility to actors inside and outside of the United States federal government.*

*Performance Information Framing as a Presentational Strategy*

Performance measurement systems are premised on the hope they can make blame attribution easier, by rendering governmental outcomes more legible to the public. But performance measures are frequently subjected to presentational strategies by elected officials who seek to muddy responsibility attribution (Bevan and Hood 2006), and bias processes of interpretation (James, Moynihan, Olsen and Van Ryzin 2020).

While how much people actively use performance information to make decisions is up for debate, a variety of studies suggests that citizens do use that information to form judgments and make decisions, especially in visible and salient service areas, such as education and health. In the United States, if schools were categorized as failing under the No Child Left Behind Act, they became less likely to win new resources via referenda (Kogan, Lavertu, and Peskowitz, 2016). Parents were also more apt to move their kids to other schools (Holbein 2016). White parents, in particular, were more likely to exit schools and to vote in local school board elections after negative performance scores, while black parents were more likely to vote but not to exit (Holbein 2019).

Such work also suggests the possibility of a negativity bias in how people use performance information in making responsibility attributions: the idea that negative performance scores gain attention and activate responsibility attribution in a way that positive performance does not. For example, James and John (2007) find that people punish incumbents for poor performance but observe no significant gains in support when an incumbent performs well. Similarly, Olsen (2015) finds that people exposed to negatively framed public health information are more likely to engage in attributional reasoning than individuals exposed to positively framed information. Such attributions may affect decisions around funding and management reforms (Nielsen and Baekgaard 2015), and even how public managers are treated: for example, elected officials show the same tendency to focus on poor performance when attributing blame to school principals (Nielsen and Moynihan 2017).

One reason for negativity bias is that the audience is actually discriminating between different levels of performance, and appropriately blaming poor performers. Another reason is that negative numbers trigger cognitive processes that focus on failure (Baumeister et al. 2001; Olsen 2015; Rozin and Rozyman 2001). To get solely at the latter process, we use equivalence framing (Belardinelli et al. 2018; Fuzenzalida et al. 2018; Olsen 2015). In other words, respondents are shown the same public health performance information but in some instances it was negatively framed (percent of public seeking a test that was not tested for the virus), and in some cases it will be positively framed (percent of the public seeking a test that was tested for the virus). The use of equivalence frames thereby limits evaluations to how the information is framed, rather than about the actual level of performance. Negatively framed performance information should elicit more negative performance evaluations than logically equivalent positively framed performance information.

Another advantage of equivalence framing is that it closely mirrors presentational strategies aimed at "spinning" performance information: when politicians can't change the actual level of performance they try to change its meaning (Bevan and Hood 2005; Moynihan 2008). In the context of COVID-19, a closely watched indicator of the government's efforts was its ability to provide tests for the virus. President Trump tried to frame testing rates positively, by insisting that many tests have been done and that their number had rapidly increased, and by repeatedly asserting the false claim that everyone who wanted a test could get one (Bump 2020). This effort to positively spin testing numbers is intended to avoid the negative reasoning that comes from looking at the same numbers and noting that the volume of tests did not meet demand. Thus, we use a widely employed frame to isolate negative reasoning that is thematically consistent with actual political efforts to positively frame one of the most closely scrutinized performance indicators of the response.

*H2: Exposure to negatively framed performance information will trigger more negative performance evaluations of actors in the United States federal government than exposure to positively framed performance information.*

*H3: Exposure to negatively framed performance information will trigger greater responsibility attribution to actors in the United States federal government than exposure to positively framed performance information.*

Public assessments of performance information are also colored by motivated reasoning. Compelling evidence on the intersection of processes of interpreting performance information and motivated reasoning comes from Van Bekerom, van der Voet, and Christensen (2020). They show that public organizations face a negativity bias in the form of greater punishment from members of the public for poor performance relative to private firms, with the effect driven by those with an ideological predisposition to believe private firms performed better. Other studies tie motivated reasoning to partisan identity. Republican voters, who generally opposed the Affordable Care Act, were more apt to select performance information that made the Act look less positive and interpret performance information about the Act more negatively, especially when primed to think about their political identity (James and Van Ryzin 2017; Jilke 2018). Bisgaard (2019: 824) shows that even though partisans may evaluate performance information accurately, they are selective in how they attribute responsibility, doing so in ways that "fit their preferred worldviews."

*H4: Conservatives exposed to the negatively framed performance information will be more likely to blame actors outside of the United States federal government when compared to those exposed to positively framed performance information.*

*Scapegoating as a Presentational Strategy*

Public management research is most attentive to directional reasoning that is driven by political ideology. However, there are other sources of directional reasoning. The search for justification for a judgment is biased by a more profound effort on the part of the evaluator to maintain a positive self-image, tied to what is sometimes referred to as the fundamental error of attribution: when performance is good, we find reasons to take credit and when performance is bad we find reasons to blame others. For example, teachers attribute responsibility to their own efforts when shown data that indicates that their school performs well. But if shown data that the school is performing poorly, they are apt to blame others, such as elected officials, the Ministry of Education, or parents (Petersen, Laumann and Jakobsen 2018).

The persistent tendency to privilege certain explanations of observed performance over others results in attribution errors, where an evaluator crafts narratives that absolve them and the groups they identify with from responsibility for negative events (Haider-Markel and Joslyn 2017: 361). These outgroups can be constructed in different ways. An obvious source of difference is ethnic or racial differences. For example, Fishman, Rattner, and Weinmann (1987) found that Israeli Jews were more likely to blame Israeli Arabs for crimes than other Israeli Jews. Ben-Porath and Shaker (2010) show that including photographs of black Hurricane Katrina victims from the City of New Orleans resulted in whites blaming (predominantly black) residents of the City more for the consequences of Hurricane Katrina. Maeder, Yamamoto, McManus, and Capaldi (2016) show that a sample of predominantly white participants were significantly more likely to blame black defendants for a crime than white defendants, despite the circumstances surrounding the crime being the same. Within public management research, racial differences are well-established as a potential source of bias in how street-level bureaucrats and policymakers evaluate minority groups, especially in the context of welfare services (Jilke and Tummers 2018). However, we have no evidence as to where ethnic scapegoat triggers fall in the blame attribution landscape when the public is asked to examine performance information.

In the context of COVID-19, the use of the term "Chinese virus" offered what appeared to be a straightforward effort to shift blame by triggering outgroup biases. The World Health Organization, who typically name such viruses, has a policy of not using names that evoke specific places or groups to avoid stigma (New York Times, 2020). Their official choice of COVID-19 was established on February 11. From early March some Republicans began to use the term Chinese virus or Wuhan Virus, and Trump included Chinese virus in a tweet by mid-March, before making it a staple – and a subject of reporter's questions – in his daily press briefings related to the pandemic. While President Trump acknowledged this presentational strategy was intended to direct attention at the culpability of the Chinese government, it raised concerns about stigmatizing Asian-Americans generally (Rogers, Jakes, and Swanson 2020). "*Spit On, Yelled At, Attacked: Chinese-Americans Fear for Their Safety*" read one headline (Tavernise and Oppel Jr. 2020).

Noting these reports, and acknowledging that public health is a public service whose performance is shaped by inputs from both government and the public, we test if being exposed to the term "Chinese virus" in a description of COVID-19 acts as a racial or nationalistic cue that shields the government from blame, and causes greater blame towards Chinese residents of the United States and the Chinese government. We also examine if this effect is greater for conservatives who identify with the political actors associated with such strategies.

*H5: Participants exposed to the term 'Chinese virus' will be more likely to blame Chinese residents and the Chinese government when compared to participants exposed to the term 'COVID - 19.'*

*H6: The effect of the "Chinese virus" framing on blaming Chinese residents and the Chinese government will be greatest for conservatives.*

**Setting and Research Design**

We use a survey experiment to examine the effect of two presentational blame avoidance strategies – scapegoating and equivalency framing – on responsibility attribution during the 2020 COVID-19 pandemic. On January 21st, 2020 the United States reported its first confirmed case of COVID-19 (CDC 2020). By March 21st, the United States reported 24,583 confirmed cases of COVID-19. Around the same time, the Trump administration began to hold daily press conferences where performance information, which often focused on testing rates, was provided to the public. During these press conferences, President Trump would also frequently refer to COVID-19 as the "Chinese virus." By late June, the United States reported over 2.5 million confirmed cases of COVID-19 and more than 120 thousand related deaths. These numbers were more than twice those of Brazil, the country with the next highest number of confirmed COVID-19 cases and related deaths. Our experiment was run between June 24[th] and June 27[th], 2020, four days after a high-profile campaign rally in Oklahoma held by President Trump where the president prominently referred to COVID-19 as the "Chinese virus" and "Kung-flu."

In our survey experiment, respondents were randomly assigned to one of four different treatment groups, representing a 2 by 2 between subjects design, plus one additional baseline group (Appendix A provides the exact wording of the prompts). The baseline group provides estimates for patterns of motivated reasoning *absent* experimental exposure to our performance framing or scapegoat cue. However, it is important to acknowledge that participants assigned to a control will have been exposed to versions of such treatments from the universal media coverage and political messaging around the pandemic. Our experimental results should therefore be read as relatively conservative tests of the blame avoidance strategies, since the non-treated subjects will, to some degree, have been exposed to those same strategies. While a limitation of our research, this is an inevitable tradeoff of trying to model highly visible and salient blame avoidance efforts.

In the baseline group, individuals are told that the Trump administration is dealing with the challenge of testing residents for a new and potentially dangerous virus - no performance information and no ethnic outgroup cue is provided. The 2 by 2 treatment matrix varies on a) whether performance information is framed positively or negatively and b) whether the term "Chinese virus" or the more neutral "COVID-19" is mentioned. Our performance information are presented as COVID-19 testing capacity, which is framed in terms of the percent of people seeking tests who can be tested (e.g., 65%) versus percent of people seeking tests who cannot be tested (e.g., 35%). Since it is possible that respondents might be more influenced by round-number integers (James, Moynihan, Olsen and Van Ryzin, 2020), thus as a form of stimuli sampling, we randomly varied the integers that subjects were exposed to (between a range of 51-99% for those who can be tested and 1-49% for those who cannot be tested).

Finally, we conducted chi-square tests to examine whether randomization was successful and all groups to which participants were assigned were balanced on key covariates. Results reveal no significant differences across treatment groups in terms of participant race ($p$ = .789), education ($p$ = 0.301), age ($p$ = 0.384), party affiliation ($p$ = 0.273), political ideology ($p$ = 0.237), political trust ($p$ = 0.294), and gender ($p$ = 0.953). We do detect an imbalance across treatment groups in terms of income ($p$ = 0.047). Therefore, we ran all models controlling for income and detected no significant variation when compared to models that did not control for income Analyses that include income as a control variable can be found in the supplementary materials.

**Sample**

A total of 1500 United States based participants were recruited for our experiment using CloudResearch, a research platform that integrates with Amazon's Mechanical Turk (MTurk)[2]. While MTurk allows requestors to recruit workers to fulfill a range of tasks, such as responding to surveys, it was not designed specifically for the purposes of conducting academic research. Noting the popularity of MTurk among researchers, the CloudResearch platform was created for the specific purpose of assisting researchers recruit MTurk participants to take part in research by, for example, allowing researchers to screen out multiple responses from the same worker and to recruit a more diverse pool of participants (Litman, Robinson, and Abberbock 2017), although online convenience samples like ours are not representative of the general population. Researchers have raised concerns over the possibility of demand effects (Mummolo and Peterson 2019) in online convenience samples, but responses appear to be highly comparable to those obtained using more representative samples (Coppock, Leeper, and Mullinix 2018; Coppock and McClellan 2019; Mullinix, Leeper, Druckman, and Freese 2015).

---

[2] An explanation of how the sample size was calculated is included in the preregistration.

Sampling criteria used in recruitment included approval ratings (greater than 94%) and number of hits (greater than 1000)[3]. Data were cleaned following procedures outlined by Mason and Suri (2012) as well as those of Dennis, Goodson, and Pearson (2018). Incomplete responses, responses completed in an unusually short amount of time (less than 180 seconds), and multiple responses from the same IP address were screened out, leaving 1439 usable responses.

## Measured Variables

We evaluate political ideology using a 6-point scale that ranges from very conservative to very liberal. Using this scale, we create a conservative dummy variable, that groups participants who identify as very conservative, conservative, and slightly conservative together as conservative (1), and participants who identify as moderate, slightly liberal, liberal, or very liberal as not conservative (0). In the supplementary materials, analyses are replicated using a Republican dummy variable.

---

[3] Approval rating is a continuous scale that runs from 0 to 100%. An approval rating of 0 indicates that the work done by this MTurk participant was negatively evaluated by the requester (i.e., researcher) for every task they performed. By contrast, an approval rating of 94% indicated that an MTurk participant, on average was positively evaluated for every task they participated in. The number of hits means how many tasks an MTurk participant has carried out.

This study uses close and open-ended dependent variables (provided in Appendix B) to assess responsibility attribution from three key perspectives: motivated reasoning, performance information framing and outgroup bias.

Our close-ended items account for the range of specific government and non-government actors participants can assign blame to. The first close-ended item asks participants to evaluate the performance of the Trump administration in responding to the pandemic on a same scale of 0 (extremely bad) to 10 (extremely good). The second asks participants how severe a threat they believe COVID-19 is according to a 0 (completely overblown) to 10 (extremely serious) scale. The third set of close-ended items asks participants how responsible the Trump administration, Obama administration, the Chinese government, Nancy Pelosi, and a range of additional actors frequently mentioned by President Trump are for the spread of COVID-19 in the United States. Response options ranged from 0 (not responsible at all) to 10 (extremely responsible). The final close-ended item evaluates responsibility attribution to different social groups, asking participants which ethnic group is most responsible for the spread of the virus across the United States. Response options include: Chinese, Hispanic, Black, White/European, and Korean. We then collapse these items into a binary variable (1 = Chinese, 0 = non Chinese).

Relying on close-ended attribution measures alone means we may miss subtle ways discriminatory attitudes that arise in response to the presentational strategies we examine. Open-ended responses "provide a direct view into the respondent's own thinking" (Roberts et al. 2017: 1065) and avoid forcing participants to interpret events through a lens of pre-constructed categories developed by the researcher. We can therefore avoid problems of demand effects and social desirability that often are of a concern for close-ended questions. Similarly, Iyengar (1996: 64) writes that open-ended questions are beneficial because they "do not cue respondents to think of particular causes or treatments." We therefore include an open-ended item that asks participants to share their opinions in 20 characters or more on who they feel has done a bad job in responding to the pandemic.

**Data Analysis Strategy**

To examine the effect of political ideology and our treatments on close-ended items, we use a series of between-subjects ANOVA. Shapiro-Wilk tests showed that none of the closed-ended dependent variables were normally distributed (for all variables: $p$ <0.001). To account for this, bias corrected and accelerated bootstrapping with 2000 replacements from the full sample was used to construct 95% confidence intervals around the mean differences (Efron 1987).

To analyze responses to open-ended items, we use an unsupervised method of text analysis referred to as structural topic modeling (STM) (Roberts et al. 2014). As a topic model, this semi-automated approach to analyzing textual data "infers" rather than "assumes" topics and the basket of words that they consist of (Roberts et al. 2017). A key feature of STM is that it allows researchers to estimate topics using text and available document metadata, which in our case would be political ideology of the respondent and the presentational strategies/manipulations they were exposed to (Roberts et al. 2014).

STM, like vanilla topic models, are mixture models which implies that any given document can consist of multiple topics. Topics can be understood as distributions of word groupings that correspond to a theme. An important goal of STM is to identify "a set of shared latent topics across a set of documents and evaluate potential relationships between document-level covariates and the prevalence of a given topic" (Bogozzi and Berliner 2016).

To determine the number of topics to extract in small corpora (i.e., "a few hundred to a few thousand" documents) we follow the recommendation of Roberts, Stewart, and Tingley (2014) and closely examine 5 to 15 topics for interpretability and redundancy. Guided by these insights and an empirical analysis of the number of topics that best balance semantic coherence of topics with exclusivity, we opted for a five-topic structural topic model. Details can be found in the supplementary materials. The results of this empirical analysis can be found in the supplementary materials.

**Results**

We present our results, first presenting findings using close-ended data and then reporting on responses to open-ended questions to offer a more comprehensive and nuanced understanding of the attribution patterns identified.

*Hypothesis 1: Effects of Motivated Reasoning*

We assess the evidence on motivated reasoning using participants assigned to our control group, who were not exposed to the performance framing or scapegoating treatments. This group therefore gives both a baseline understanding of how conservatives differ from others in how they view the crisis, and provides a sense of the context against which our experimental treatments must be considered.

*H1: Close-ended analysis*

The general pattern of results for close-ended items reveals enormous differences in how conservatives and non-conservatives evaluate government performance around the same event, offering clear support for Hypothesis 1. In response to the question, "How serious of a threat do you think the virus you just read about is? (0 = Completely overblown, 10 = Extremely Serious)" conservatives viewed the threat as less serious (see Figure 1), rating it 5.9, compared to 7.7 for moderates and liberals ($p < 0.001$). Republicans were also more generous in evaluating "the leadership in the United States in responding to this pandemic." On a 0-10 scale, where 10 is extremely good, conservatives averaged 6.3, while moderates and liberals averaged 2.2 (see Figure 1).

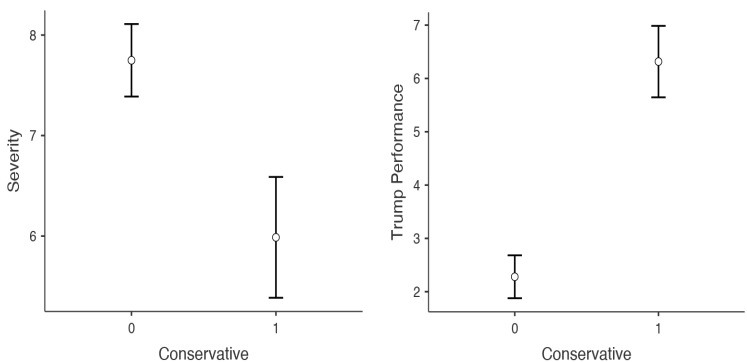

Figure 1: Perceived severity of COVID-19 and Trump administration performance evaluations among conservatives assigned to the baseline group. Vertical lines represent 95% confidence intervals.

While conservatives are more supportive of political leadership, 6.3 is not a very high score, reflecting some potential misgivings with performance. At the same time, conservatives appear more willing to shift blame away from Trump and his administration. In a series of questions where we asked which groups were most responsible for the spread of the virus, conservatives differ from other respondents. We start with attributions of responsibility toward highly polarized actors that President Trump has frequently targeted: former President Obama and the mainstream media (Figure 2). Given Obama has been out of office since the beginning of 2017, and the media has no role beyond covering the crisis, judging them to be responsible for the spread of the disease requires significant motivated reasoning. Asked how responsible Obama was for the spread of COVID-19 in the United States, the mean among conservatives was 2.5 and for moderates and liberals, 1.1 ($p < 0.001$). For mass media, the mean among conservatives was 5.1 and for moderates and liberals 3.5 ($p < 0.001$). Additionally, conservatives were also more inclined to blame Speaker Pelosi and the Chinese government than moderates and liberals. The mean among conservatives for Pelosi was 3.9 and for moderates and liberals, 2.1 ($p < 0.001$). With respect to blaming the Chinese government, the mean among conservatives was 7.8 and among moderates and liberals, 5.8 ($p < 0.001$). By contrast, conservatives were less inclined to assign blame to President Trump – the mean among conservatives was 4.1 and among moderates and liberals, 7.3 ($p < 0.001$). Conservatives were also more inclined to blame the CDC with the mean among conservatives being 5.2 and among moderates and liberals 4.1 ($p < 0.001$).

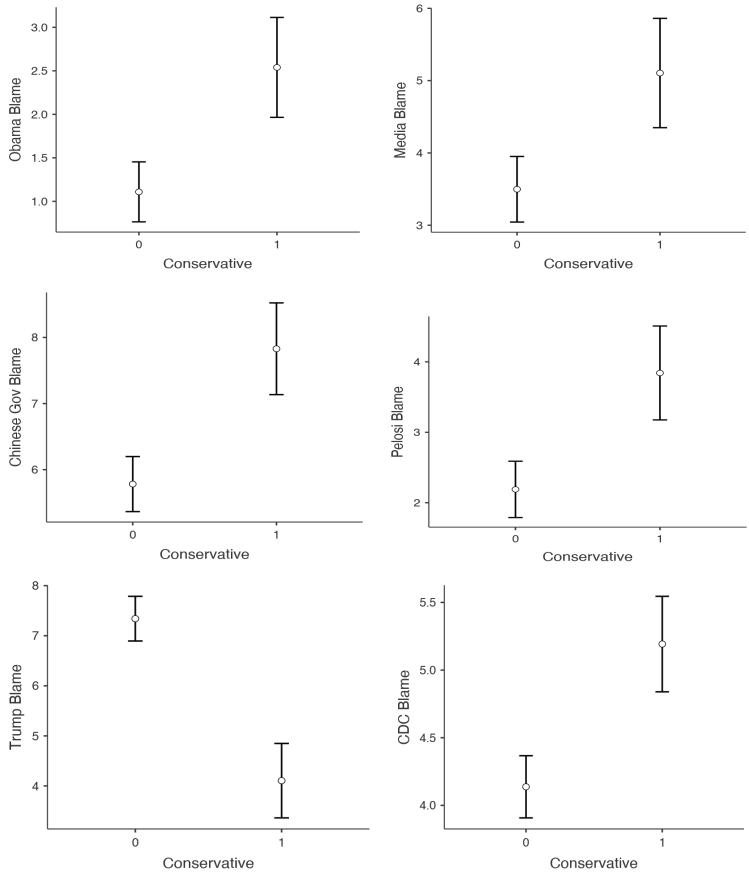

Figure 2: Effect of partisan motivated reasoning on blame attribution among participants assigned to the baseline group. Vertical lines represent 95% confidence intervals.

Conservatives are also more apt to blame Chinese residents when provided a list of ethnic groups in the United States, and asked which one is the most responsible for the spread of the virus. To put the responses in perspective, Asian-Americans as an entire group make up less than six percent of Americans, while whites represent more than three-quarters. There is no evidence that COVID-19 is more prevalent among Asian-Americans: one analysis found that they represented less than five percent of COVID-19 deaths (APM Research Labs 2020), while further work indicated that the New York COVID-19 outbreak had originated in Europe (Zimmer 2020). Based simply on the population size differences, the most plausible response is that whites are most responsible for the spread of the disease. Nevertheless, 62% of conservatives identified Chinese-Americans as the ethnic group most responsible for the spread of the disease, compared to 30% of non-conservatives. Interestingly, conservatives were also less likely to blame whites for the spread of COVID-19 in the United States when compared to moderates and liberals – just 34% of conservatives blamed whites compared to 63% of moderates and liberals. The large differences between conservatives and non-conservatives in attributing blame to ethnic groups implies a potential ceiling effect in our experimental treatments, since the non-treated already appears to have internalized attribution patterns consistent with the President's use of the term "Chinese virus."

*H1: Open-ended analysis*

The open-ended data provides more evidence of motivated reasoning. In response to the question of "who did a bad job in responding to the pandemic," President Trump is central to the top two topics. The remaining three topics cite states' responses and initial issues related to confusion and availability of masks. President Trump appears frequently in these topics as well. This information is presented in Figure 3.

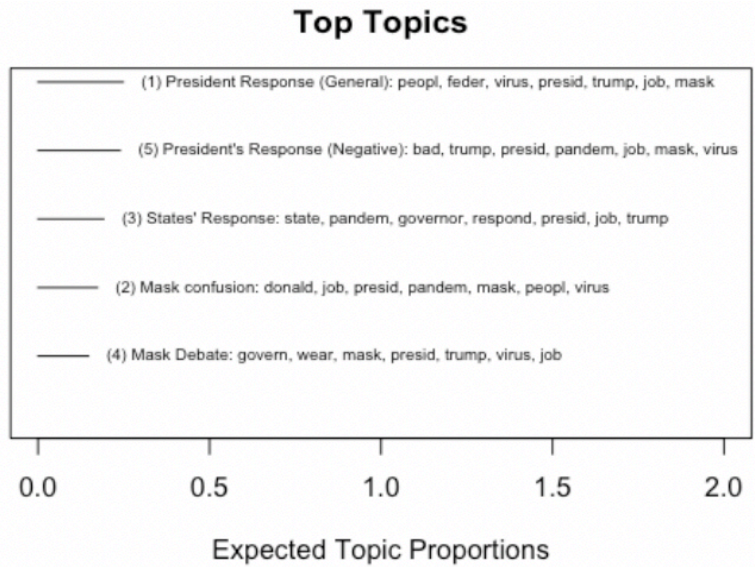

Figure 3: 5 most frequent topic proportions among all participants in the control group for the question *Who do you think did a bad job responding to the pandemic? Topic prevalence is estimated using the conservative variable.*

Figure 4 shows the impact of identifying as a conservative on the chances of a participant discussing a particular topic. As with the close-ended items, we observe that conservatives are less likely to blame President Trump and more likely to blame other actors and causes – namely, states' responses and confusion surrounding mask wearing guidelines and availability. The open-ended responses have the effect of removing constraints on the range of actors participants can attribute responsibility to as the close-ended responses did. Even when doing so, we find no evidence of consensus in the way participants from different ideological viewpoints assigned blame. In every single one of the five topics most frequently occurring topics, emphasis varied significantly according to whether a participant identified as conservative or not.

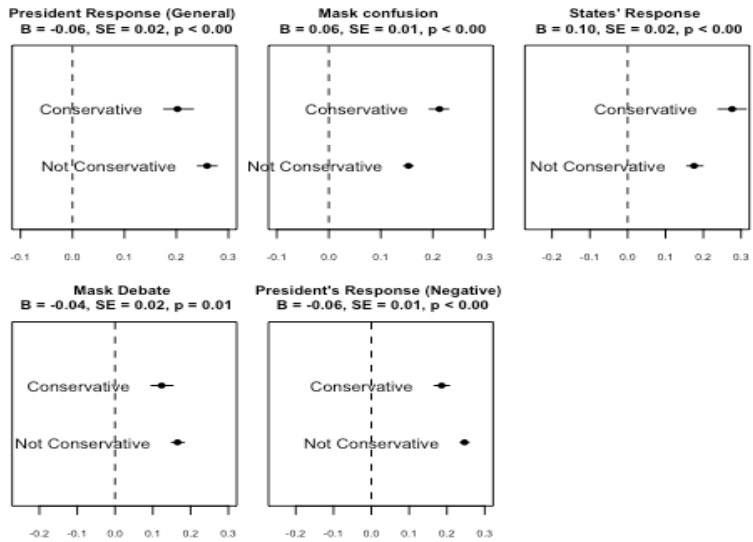

Figure 4: Graphical display of topical prevalence contrast between conservatives and non-conservatives in the control group for the question *Who do you think did a bad job responding to the pandemic?* Horizontal lines represent 95% confidence intervals.

*Hypotheses 2 and 3: Performance Information Framing as a Presentation Strategy*

### *H2 and 3: Close-ended analysis*

Hypotheses 2 and 3 predicted that performance evaluations of and responsibility attribution to the actors in the federal government (e.g., President Trump or the CDC) would differ depending on whether participants were assigned to positively or negatively framed performance information. Here, we focus on participants who were exposed to treatment. These hypotheses do not consider political ideology. The effects of performance information framing on evaluations of the Trump administration's performance during the pandemic is above standard levels of statistical significance ($F(1, 1148) = .48$, $p = 0.49$). Additionally, we find no evidence that exposure to performance information framing impacted responsibility attribution to the Trump administration ($F(1, 1148) = .004$, $p = 0.94$) or federal agencies charged with responding to the pandemic (i.e., the CDC) ($F(1, 1148) = 1.12$, $p = 0.29$). These findings are illustrated in Figure 5 below.

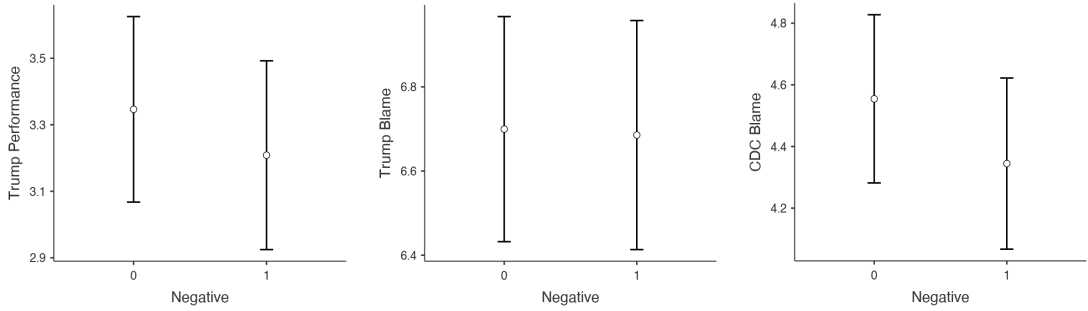

Figure 5: The impact of performance information framing on performance evaluations of Trump administration Performance and Responsibility Attribution to Trump administration and CDC for the spread of COVID-19 in the United States for the full sample of participants. Vertical lines represent 95% confidence intervals.

### *H2 and H3: Open-ended analysis*

The open-ended responses provide a slightly different perspective on how exposure to performance information framing impacts blame attribution. Results of the five most frequent topics, shown below in Figure 6, point to a diverse set of actors that appears to split along partisan lines – the two most frequent topics related to the federal government's response and President Trump, and the remaining three highlight issues with Democrats' role in the economic downturn, masks, and state governments.

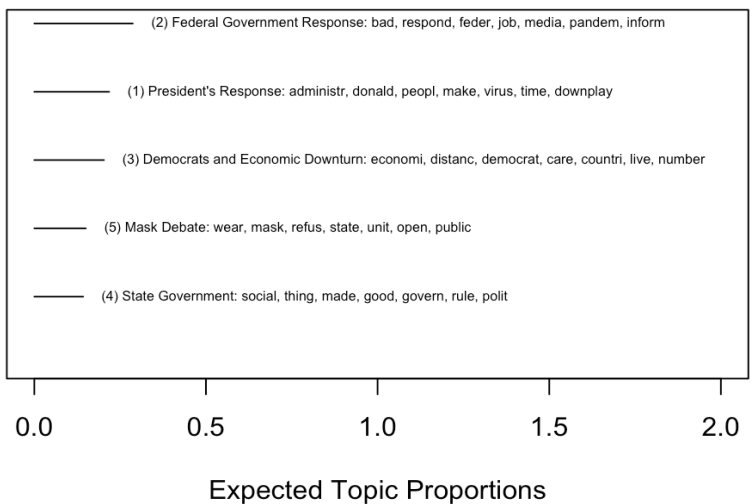

Figure 6: 5 most frequent topic proportions among all participants for the question *Who do you think did a bad job responding to the pandemic?* Topic prevalence is estimated using the performance information framing variable.

Figure 7 shows the impact of exposure to negatively versus positively framed performance information on the frequency with which a particular topic is discussed. In contrast to the close-ended items, we see that the frequency with which a topic is discussed varies significantly according to whether participants were exposed to negatively or positively framed performance information for two key topics – blame assigned to the federal government and to Democrats' role in the economic downturn. Specifically, when compared to participants exposed to positively framed performance information, those exposed to negatively framed performance information were significantly *more* likely to blame Democrats for their role in the economic downturn, and significantly *less* likely to assign blame to the federal government's response to the pandemic. Blaming Democrats for the economic downturn may reflect decisions over whether to open or close the economy in response to the pandemic, with Democrats generally advocating for closures and Republicans advocating to remain open. While we observe a significant impact of performance information framing on blame attribution for the federal government, there is no significant impact on blame attribution to the president. This is interesting in that it alludes to a level of nuance in the way exposure to negatively framing performance information, as a presentation strategy, impacts responsibility attribution.

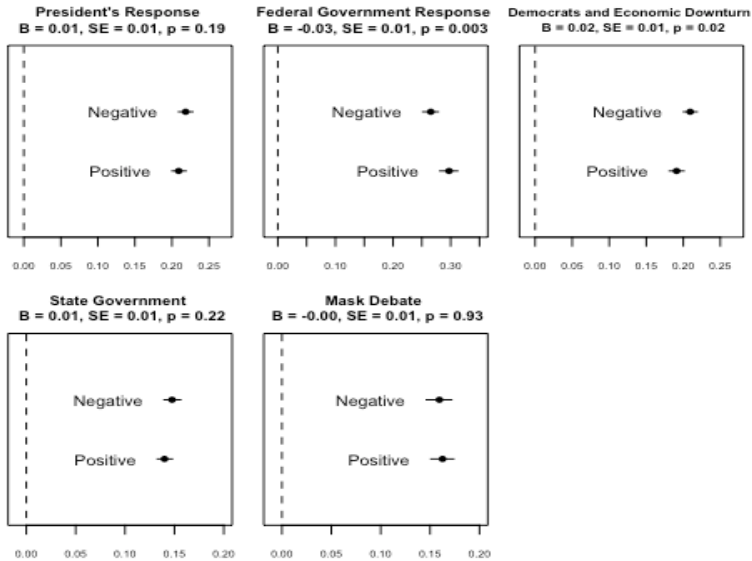

Figure 7: Graphical display of topical prevalence contrast between exposure to positively and negatively framed performance information among all participants for responses to the question *Who do you think did a bad job responding to the pandemic?* Horizontal lines represent 95% confidence intervals.

## Hypothesis 4: Partisan Motivated Reasoning and Performance Information Framing

### H4: Close-ended analysis

Hypothesis 4 predicted that conservatives exposed to the negatively framed performance information would be more inclined to blame actors outside of the United States federal government when compared to those exposed to positively framed performance information. In other words, exposure to negatively framed performance information would trigger blame shifting among conservatives. Responses to close-ended items do not offer support for this prediction. Focusing on frequent targets of the Trump administration, we find no significant difference among conservative participants assigned to positively and negatively framed performance information in their responsibility attributions to former President Obama (F(1, 341) = 1.80, $p$ = 0.18), Congresswoman Pelosi (F(1, 341) = 2.86, $p$ = 0.09), the Chinese government (F(1, 341) =

1.13, *p* = 0.28), and mass media (F(1, 341) = 0.920, *p* = 0.33). Differences across treatment groups are illustrated below in Figure 8.

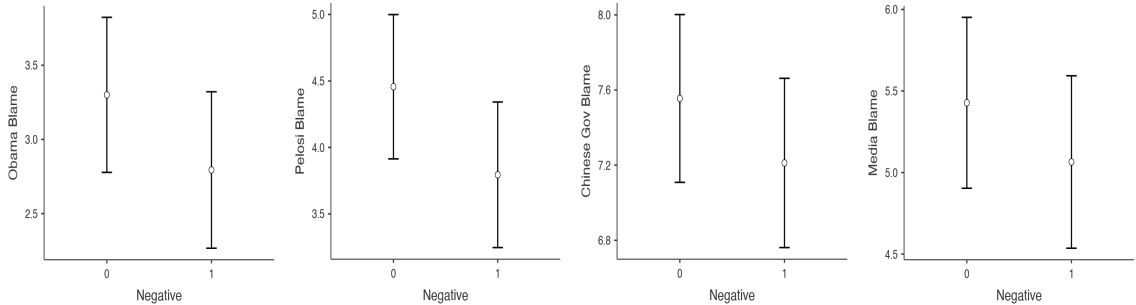

Figure 8: The impact of performance information framing on responsibility attribution to the Obama Administration, Speaker Pelosi, the Chinese government, and the mass media for the spread of COVID-19 in the United States among conservatives. Vertical lines represent 95% confidence intervals.

### *H4: Open-ended analysis*

For conservative participants, responses to open-ended responses provide a broader view on the impact of performance information framing on blame attribution.

The five most frequently mentioned topics, shown in Figure 9, among conservatives,

clearly echo President Trump's speaking points. The three topics participants most

frequently assigned blame to were Democrats, governors, and the "liberal left."

Following this, the remaining two topics, which appear much less frequently than the top three topics, include the President's response and a generic 'administration' category.

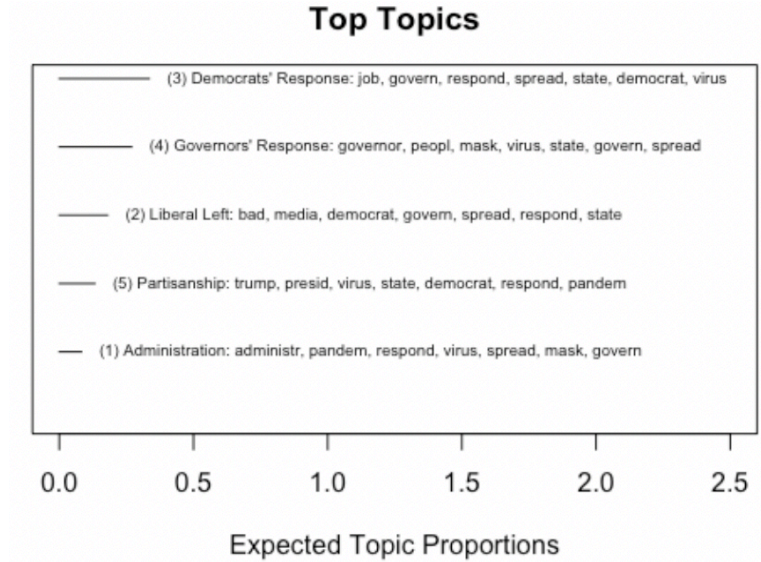

Figure 9: Most frequent topic proportions among conservatives for the question *Who do you think did a bad job responding to the pandemic?* In estimating these topics, prevalence is estimated using the performance information framing variable.

Figure 10 illustrates how topic prevalence varies according to performance information framing among conservative participants. Here again we observe significant variation that results from exposure to performance information framing. However, this variation demonstrates that exposure to negatively framed performance information did not have a standard effect, but relatively nuanced attributional reasoning. When given negative information, conservatives do not become more likely to blame the Trump administration, as with the close-ended responses. But they do become more likely to seek to blame, and their ire is directed toward a diverse bipartisan set of actors – state governors – and the abstract concept of partisanship. Thus, this presentational strategy appears to shield the Trump administration from the blame arising from bad news. When provided a positive frame, conservatives are more willing to blame partisan targets of President Trump's criticisms - Democrats and the liberal left. Here, it may be that the good news of the positive frame causes conservatives to punish actors they believe have been too critical of the Trump administration. Cumulatively, these findings offer some support for Hypothesis 4, with the qualification that exposure to framing operates in more nuanced ways than established by the close-ended categories. Frames did increase blame shifting by conservatives, in a way that generally protected the Trump administration, even as positive and negative frames were processed differently to assign blame to different actors.

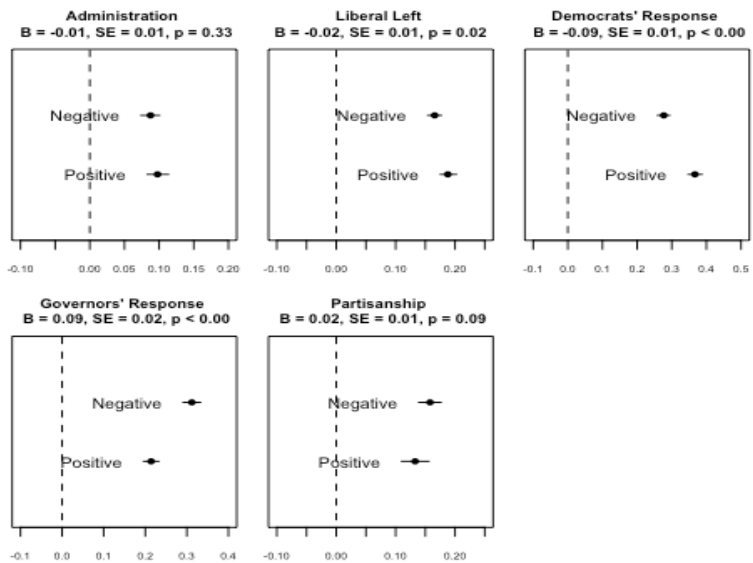

Figure 10: Graphical display of topical prevalence contrast between exposure to positively and negatively framed performance information among conservatives for responses to the question *Who do you think did a bad job responding to the pandemic?* Horizontal lines represent 95% confidence intervals.

*Hypothesis 5: Effects of Scapegoating as a Presentational Strategy*

*H5: Close-ended analysis*

We first examine the impact of the scapegoat trigger on assigning blame for the spread of COVID-19 in the United States to the Chinese government and Chinese residents for the entire sample (H5). Participants exposed to the scapegoat cue – Chinese virus – when communicating performance information were more inclined to blame Chinese residents for the spread of COVID-19 in the United States when compared to those who were not exposed to the scapegoat cue (i.e., the term COVID-19) (F(1, 1150) =

6.39, $p = 0.012$). Conversely, there was no difference in blame assigned to the Chinese government across treatment groups (F(1, 1150) = 0.305, $p = 0.58$). These results are illustrated in Figure 11.

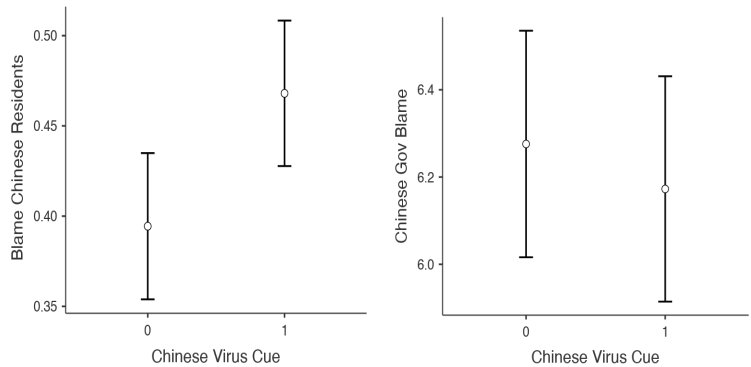

Figure 11: The impact of racial scapegoat trigger on assigning blame for the spread of COVID-19 in the United States to Chinese residents and the Chinese government for the full sample. Vertical lines represent 95% confidence intervals.

### H5: Open-ended analysis

Open-ended responses again broaden our understanding of how people respond to scapegoating. Here, we find that the scapegoating presentational strategy directs blame towards the person who pushed the term: President Trump and the federal government. The remaining three most prevalent topics center on Democrats and the economic downturn, mask debates, and a more generic category blaming government in general. The topics, along with the prevalence are reported in Figure 12.

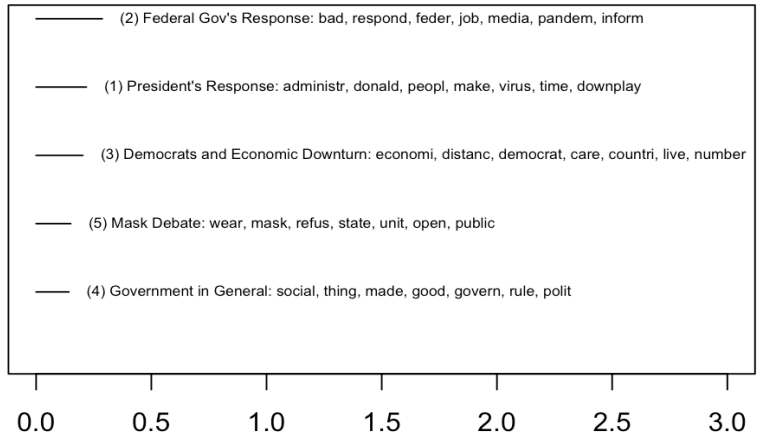

**Top Topics**

- (2) Federal Gov's Response: bad, respond, feder, job, media, pandem, inform
- (1) President's Response: administr, donald, peopl, make, virus, time, downplay
- (3) Democrats and Economic Downturn: economi, distanc, democrat, care, countri, live, number
- (5) Mask Debate: wear, mask, refus, state, unit, open, public
- (4) Government in General: social, thing, made, good, govern, rule, polit

0.0   0.5   1.0   1.5   2.0   2.5   3.0

Expected Topic Proportions

Figure 12: 5 most frequent topic proportions among participants assigned to the Chinese virus and COVID-19 treatment for the question *Who do you think did a bad job responding to the pandemic?* In estimating topics, prevalence is estimated using the scapegoat treatment variable.

Additionally, we observe variation in the prevalence of these topics attributable to exposure to the scapegoat cue. Specifically, as illustrated in Figure 13, participants exposed to the term "Chinese virus" were significantly more likely to direct attention to President Trump. Conversely, this group was less likely to discuss the mask debate than participants exposed to the term COVID-19. Interestingly, while we find an effect of the cue on blaming President Trump, we do not find any impact on blaming the federal government. This suggests that the use of the term "Chinese virus," and consequent resentment, is more closely linked to President Trump than his administration.

It appears that, when prompted by response options in close-ended items, participants will cite Chinese residents when attributing blame. However, when this prompt is absent, exposure to the scapegoat cue tends to focus the assignment of blame on the President. Of note, the term 'Chinese' was not included in any of the 5 most prevalent topics. One way of interpreting these findings is that the use of the term "Chinese virus" creates a situation where participants not only blame Chinese residents for the spread of the virus in the United States, but also the actor(s) that make use of the term.

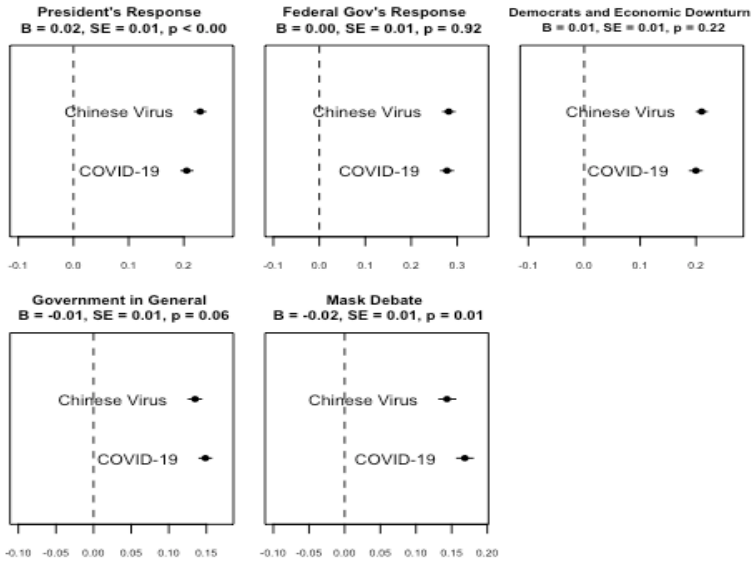

Figure 13: Graphical display of topical prevalence contrast between participants exposed to the term Chinese virus and COVID-19 for the question *Who do you think did a bad job responding to the pandemic?* Horizontal lines represent 95% confidence intervals.

*Hypothesis 6: Effects of Scapegoat Cue and Partisan Motivated Reasoning*

*H6: Close-ended analysis*

For conservative respondents we find no main effect of exposure to the term "Chinese virus" on blame assigned to Chinese residents ($F(1, 341) = 0.1.63$, $p = 0.202$) or the Chinese government ($F(1, 341) = 0.045$, $p = 0.83$). Additionally, we find the effect of the scapegoat cue does not vary significantly according to political ideology for blame assigned to Chinese residents ($F(1, 1148) = 0.002$, $p = 0.96$) or the Chinese government ($F(1, 1148) = 0.506$, $p = 0.47$). The interaction effects are illustrated in Figure 14 below. This may simply reflect the fact that, as noted above, conservatives already have a higher base propensity to blame Chinese residents, perhaps because they have internalized the repeated framing efforts of the Trump administration, or because conservatives hold a more critical view towards outgroups and immigrants generally. We therefore probe the responses of conservatives to the scapegoat trigger in greater detail below, using both close-ended and open-ended responses.

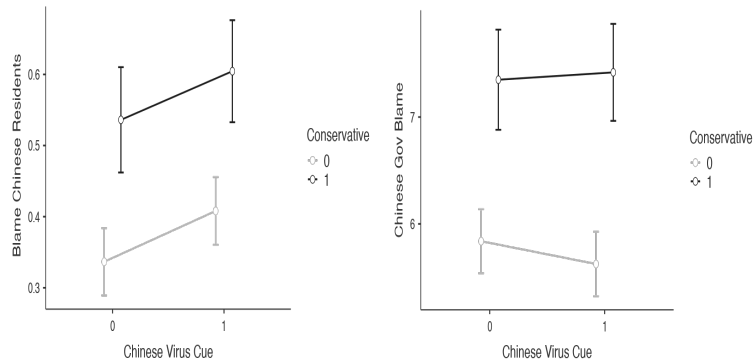

Figure 14: The impact of racial scapegoat trigger on assigning blame for the spread of COVID-19 in the United States to Chinese residents and the Chinese government for conservatives. Vertical lines represent 95% confidence intervals.

If the scapegoat cue does not increase blame to Chinese residents for conservatives, might it affect other forms of outgroup bias and blame among conservatives? We find some evidence of this, with conservatives becoming more likely to attribute responsibility to former President Obama and Speaker Pelosi when exposed to the Chinese virus cue (for Obama: $F(1, 1148) = 8.25$, $p = 0.004$; For Pelosi: $F(1, 1148) = 7.77$, $p = 0.005$). Figure 15 demonstrates these differences and also shows that exposure to the scapegoat trigger reduces assignment of blame to Obama and Pelosi among moderates and liberals.

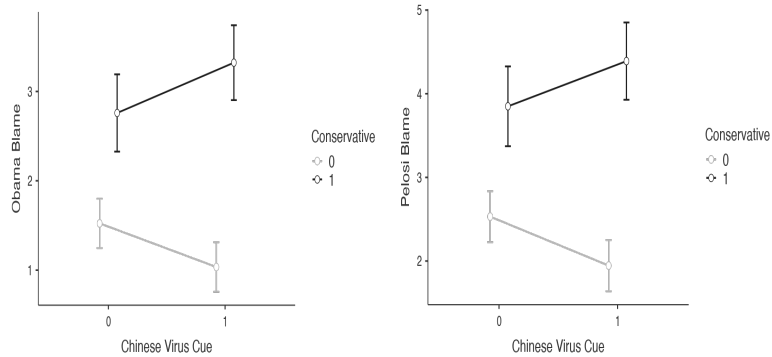

Figure 15: The impact of racial scapegoat trigger on assigning blame for the spread of COVID-19 in the United States to Obama and Pelosi according to whether a participant is conservative or not. Horizontal lines represent 95% confidence intervals.

### H6: Open-ended analysis

The open-ended responses provide additional nuance to our understanding of how conservatives respond to the Chinese Virus cue. Figure 16 provides a five-topic structural topic model for conservatives exposed to the trigger. The most frequent topic was 'everyone but the President' – participants assign blame to a number of actors, including the media, the public, and state government, but fail to specifically mention actors in the federal government, such as President Trump or the CDC. Following this, in terms of decreasing frequency, participants blamed states' responses, President Trump's response, government in general, and the President's leadership. Of note is that two of the five topic models directly address concerns over President Trump's handling of the pandemic – a more personalized category calling out President Trump in particular, and a second the emphasizes the Office of the President and his administration. While topic prevelance indicates Trump does not figure prominently in the list of actors that conservatives blame for the United States' response to the pandemic, neither is he absolved from blame. A further observation is that, while President Trump frequently attempted to shift blame to prominent Democrats and China, they are not mentioned at all in topics.

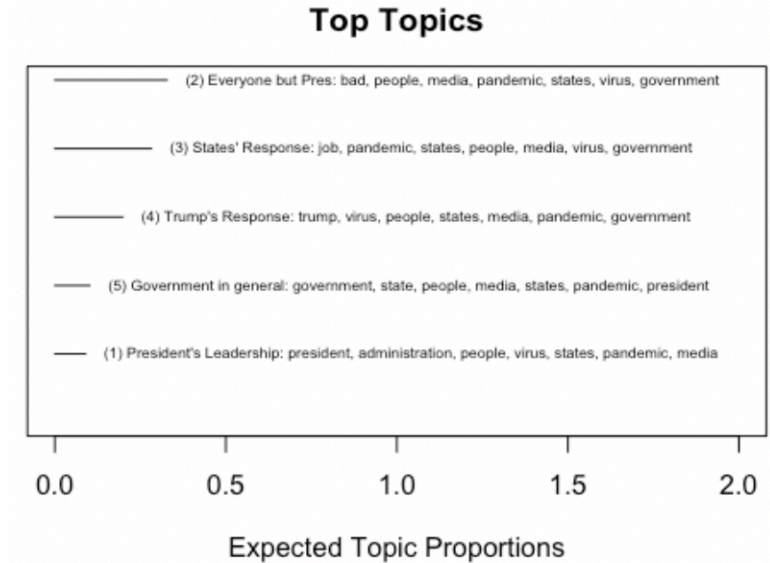

Figure 16: 5 most frequent topic proportions among conservative participants assigned to the Chinese virus and COVID-19 treatment for the question *Who do you think did a bad job responding to the pandemic?* In estimating these topics, prevalence is estimated using the scapegoat variable.

The idea that conservative participants do not completely absolve President Trump from blame comports with variation observed in the prevalence of topics attributable to exposure to the scapegoat cue. Here, we find some evidence of a backlash. Specifically, as demonstrated in Figure 17, exposing conservative participants to the term "Chinese virus" actually increased blame for Trump's response to the pandemic, and reduced the assignment of blame to state governments.

Taken together, the evidence suggests that the effect of a scapegoat cue as a presentation strategy was mixed for President Trump. It may partially deflect blame to Democrats, but does not increase blame toward Chinese residents or the Chinese government, and also actually invites it for Trump. One potential reason for this is that the term is so closely associated with Trump that merely mentioning it creates an association in the minds of subjects between Trump and the pandemic. Another possibility is that conservatives are more sensitive to international threats, and that the term "Chinese virus" triggers them to blame federal actors – not just Trump but also Pelosi and Obama – for failing to protect the borders, a failing not viewed as the responsibility of state officials.

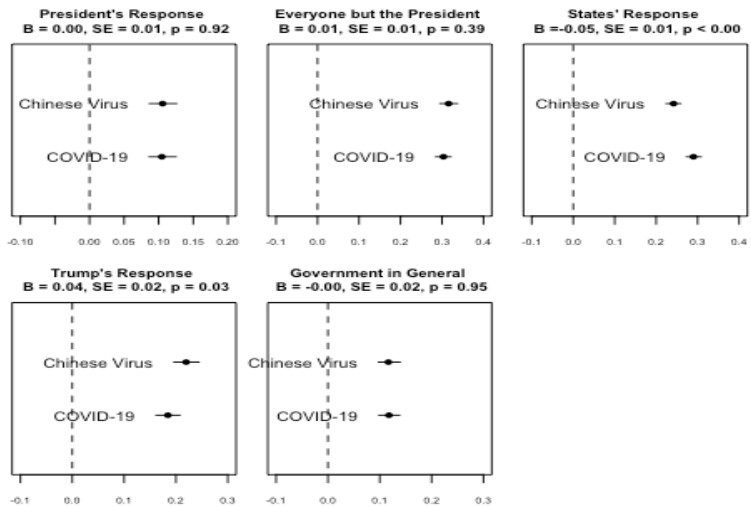

Figure 17: Graphical display of topical prevalence contrast between conservatives exposed to the term Chinese virus and COVID-19 for the question *Who do you think did a bad job responding to the pandemic?* Horizontal lines represent 95% confidence intervals.

**Discussion**

The results of the survey experiment and structural topic models provide some clear findings. We see large differences between conservatives and others in how they interpret the crisis. Conservatives are less likely to see a pandemic that arose under a Republican President as a serious threat, more likely to believe the Trump administration is doing a good job, and to blame the targets of President Trump's blame-shifting efforts: current and former Democratic politicians, the media, and the Chinese government. The results are a powerful example of partisan motivated reasoning in shaping how the public understands government performance and allocates blame as a result. The open-ended responses suggest some agreement about the cast of actors that non-conservatives and conservatives blame for the response to the pandemic, highlighting President Trump and state governments. Thus, a key point of distinction lies in the degree of blame placed on the different actors.

Second, we find evidence that the use of scapegoat cues matters. While the intent of the term "Chinese virus" is ambiguous – is it a racial cue or nationalistic cue? – the effects in our sample are not: Chinese residents face additional blame, but the Chinese government does not. While the intent of the Trump administration was to redirect blame towards the Chinese government, our findings suggest this effort failed. It is important to note here that the estimates of the scapegoat treatment are likely conservative: since the term was in frequent use by the President, many had already been exposed to it, and had already internalized the effect. This makes the increased blame assigned to East Asian and in particular Chinese Americans as a result of the use of the term all the more worrying. Invoking scapegoat triggers in a diverse society has a predictable effect of making some members of that society the target for blame. Apart from the basic need for cross-group tolerance in a democracy, the scapegoating of one group undermines the sense of collective effort needed for the public to co-produce the collective action needed to defeat a pandemic. Our results also suggest that while scapegoat cues might trigger blame toward those groups, it does not necessarily follow that the blame allocated to government declines as a result. When using the full sample, and when focusing on conservatives, our findings show that the use of such divisive language causes unease, and even increases blaming toward Trump. As an exercise in blame avoidance, the use of scapegoat triggers is risky: they create blame toward the outgroup, but may also do the same for the blamer.

Our evidence that efforts to spin performance information using equivalence framing made a difference to the judgments of our respondents is mixed. Close-ended responses, provide no evidence of a significant effect of performance information framing on blaming the specific actors and institutions listed in our response options. Equivalence testing allows us to accept the null hypothesis of no difference in performance evaluations between the negative and positive performance information treatment groups for hypotheses 2 and 3.

On the other hand, findings from the open-ended responses demonstrate a significant effect of performance information framing on blame attribution that close-ended items were unable to capture. Analyses using the full sample, and a subsample of conservatives indicate participants use negatively framed performance information to punish some actors, and positively framed performance information to blame others, and these patterns depend upon how motivated reasoning aligns those actors as enemies of the President (blamed when the framing is positive) or more abstract and bipartisan targets (blamed when the framing is negative). This finding is novel in that it extends prior work showing a negativity bias in the way individuals respond to performance information (James and Mosely 2014; Nielsen and Moynihan 2017; Olsen 2015; Van Bekerom, van der Voet, and Christensen 2020). Our findings extend this line of work, showing that participants appear to respond to both negatively and positively framed performance information, although in different ways. From this perspective, responses to performance information may be more sophisticated than previous research has implied.

**Conclusion**

Highly visible and salient crises have an outsize effect on how the public judges the quality of government. While public actors can try to shape that evaluation through skillful management of the event itself, they also devote significant attention to managing perceptions of the event by employing different presentational strategies (Hood 2010; Weaver 1986). Our findings suggest the limited power political leaders have, and significant risks they face, during an enormous crisis. Conditions of polarization may partially shield public actors, but also ensure some blame, and their own presentational blame avoidance strategies may backfire or be rendered ineffective.

An obvious limitation of our analysis is that it examines only perceptions of blame attribution in one country during a global pandemic. The particular conditions of the US setting – a generally poor response, intense polarization, and a President engaged in intense blame avoidance strategies – make it both ideal to analyze the variables we study, but also mark it as distinct from other countries. An obvious extension of our findings would be to examine how the public evaluates government performance when those conditions are less present.

One theoretical implication is that our findings help to build upon existing efforts to understand the implications of equivalence framing interacts with motivated reasoning to shape responses to performance information. The non-significant effects on close-ended responses may speak to boundary conditions for the role of equivalence framing in evaluating performance information, and suggest it may have a less marked impact in situations where respondents have strong political priors about the topic (see also Damgaard and Nielsen 2019). But we also show that standard analyses that use only close-ended indicators may miss other effects of the framing of performance information.

Just as prior work observes motivated reasoning in the processing of performance information (e.g., Van den Bekerom, van der Voet, and Christensen 2020), we extend this work, focusing on the effects of performance framing in a highly political context. The open-ended responses show respondents interpret the framing of information in creative ways that largely protects the partisan official they identify with. As the setting we focus on is unique, future research can extend these findings by examining whether a similar pattern of effects translates to other highly politized contexts. Future work could therefore do more to vary the conditions under which equivalence and other sorts of performance information framing is provided, using more realistic scenarios to gauge the strength of the frame relative to other factors such as user beliefs or factors such as political knowledge.

A final theoretical implication of our findings relates to the impact of scapegoat cues, such as the term Chinese Virus, on blame attribution. While research on blame attribution in public management is not new, extant research on determinants has largely been limited to partisanship. Noting the role of outgroup bias in shaping the assignment of blame, we extend this body of research by illustrating how, in the context of co-produced public goods, outgroup cues can trigger scapegoating of minority groups. However, one limitation of this strategy is that is focuses on a very idiosyncratic context – the last pandemic to hit the United States was in 1968 (CDC 2020). Future research could extend these findings by exploring how the use of similar scapegoat cues impact blame attribution in common public service settings, such as education.

**References**

Anastasopoulos, L.J., and A.B. Whitford. (2019). Machine learning for public administration research, with application to organizational reputation. *Journal of Public Administration Research and Theory*, *29*(3), 491-510.

APM Research Labs. (2020). The Color of Coronavirus: COVID-19 Deaths by Race and Ethnicity in the U.S. https://www.apmresearchlab.org/covid/deaths-by-race#asian

Bagozzi, B. E., & Berliner, D. (2018). The politics of scrutiny in human rights monitoring: evidence from structural topic models of US State Department human rights reports. *Political Science Research and Methods*, *6*(4), 661-677.

Baker, M. (2020). When Did the Coronavirus Arrive in the U.S.? Here's a Review of the Evidence. *New York Times*. *https://www.nytimes.com/2020/05/15/us/coronavirus-first-case-snohomish-antibodies.html*

Baumeister, R. F., Bratslavsky, E., Finkenauer, C., & Vohs, K. D. (2001). Bad is stronger than good. *Review of general psychology*, *5*(4), 323-370.

Belardinelli, P., N. Bellé, M. Sicilia, and I. Steccolini. (2018). Framing effects under different uses of performance information: An experimental study on public managers. *Public Administration Review*, *78*(6), 841-851.

Ben-Porath, E. N., & Shaker, L. K. (2010). News images, race, and attribution in the wake of Hurricane Katrina. *Journal of Communication*, *60*(3), 466-490.

Bevan, G., & Hood. C. (2006). What's measured is what matters: targets and gaming in the English public health care system. *Public Administration*, 517-538.

Bisgaard, M. (2015). Bias will find a way: Economic perceptions, attributions of blame, and partisan-motivated reasoning during crisis. *The Journal of Politics*, *77*(3), 849-860.

Bump, P. (2020). The reality of coronavirus testing continues to differ from Trump's claims. *Washington Post*. https://www.washingtonpost.com/politics/2020/04/13/reality-coronavirus-testing-continues-differ-trumps-claims/

Centers for Disease Control. (2020). Past Pandemics. Accessed July 3rd, 2020. https://www.cdc.gov/flu/pandemic-resources/basics/past-pandemics.html

Coppock, A., & McClellan, O. A. (2019). Validating the demographic, political, psychological, and experimental results obtained from a new source of online survey respondents. *Research & Politics*, *6*(1), 2053168018822174.

Damgaard, P.R., and P. Nielsen. (2019). Does Performance Disclosure Affect User Satisfaction, Voice and Exit? Survey-Experimental Evidence from Actual Service Users. *Paper Presented at the European Group of Public Administration*.

Efron, B. (1987). Better bootstrap confidence intervals. *Journal of the American statistical Association*, *82*(397), 171-185.

Fishman, G., Rattner, A., & Weimann, G. (1987). The effect of ethnicity on crime attribution. *Criminology*, *25*(3), 507-524.

Fuenzalida, J., G.G. Van Ryzin, and A.L. Olsen. (2019). Are managers susceptible to framing effects? An experimental study of professional judgment of performance metrics. *International Public Management Journal*, 1-16.

Gerber, A. and D.P. Green. (1998). Rational learning and partisan attitudes. *American Journal of Political Science*, *42*(3), 794-818.

Gilbert, D. T., & Malone, P. S. (1995). The correspondence bias. *Psychological bulletin*, *117*(1), 21.

Joslyn, M. R., & Haider-Markel, D. P. (2017). Gun Ownership and self-serving attributions for mass shooting tragedies. *Social Science Quarterly*, *98*(2), 429-442.

Heider, F. (1958). The psychology of interpersonal relations. New York: Wiley.

Holbein, J. (2016). Left behind? Citizen responsiveness to government performance information. *American Political Science Review*, *110*(2), 353-368.

Holbein, J.B. and H. J. Hassell. (2018). When Your Group Fails: The Effect of Race-Based Performance Signals on Citizen Voice and Exit. *Journal of Public Administration Research and Theory*, *29*(2), 268-286.

Hood, C. (2010). *The Blame Game: Spin, Bureaucracy, and Self-Preservation in Government*. Princeton, NJ: Princeton University Press.

Isenstadt, A. (2020). GOP Memo Urges Anti-China Assault Over Coronavirus. *Politico*. https://www.politico.com/news/2020/04/24/gop-memo-anti-china-coronavirus-207244

Iyengar, S. (1996). Framing Responsibility for Political Issues. *The Annals of the American Academy of Political and Social Science*, *546*, 59-70

James, O., and Moseley, A. (2014). Does performance information about public services affect citizens' perceptions, satisfaction, and voice behaviour? Field experiments with absolute and relative performance information. *Public Administration*, *92*(2), 493–511.

James, O., and G.G. Van Ryzin. (2017b). Motivated reasoning about public performance: An experimental study of how citizens judge the affordable care act. *Journal of Public Administration Research and Theory*, *27*(1), 197-209.

James, O., S. Jilke, C. Petersen, and S. Van de Walle. (2016). Citizens' Blame of Politicians for Public Service Failure: Experimental Evidence about Blame Reduction through Delegation and Contracting. *Public Administration Review*, *76*(1), 83-93.

James, O., A.L. Olsen, D. Moynihan, and G.G. Van Ryzin, (2020). *Behavioral Public Performance: How People Make Sense of Government Metrics (Elements in Public and Nonprofit Administration)*. Cambridge University Press.

Jilke, S. (2018). Citizen satisfaction under changing political leadership: The role of partisan motivated reasoning. *Governance*, *31*(3), 515-533.

Jilke, S., and M. Bækgaard. (2020). The Political Psychology of Citizen Satisfaction: Does Responsibility Attribution Matter? *Journal of Public Administration Research and Theory*, in press.

Jilke, S., & Tummers, L. (2018). Which clients are deserving of help? A theoretical model and experimental test. *Journal of Public Administration Research and Theory*, *28*(2), 226-238.

Jones, E. E., & Davis, K. E. (1965). From acts to dispositions the attribution process in person perception. In *Advances in experimental social psychology* (Vol. 2, pp. 219-266). Academic Press.

Kandil, C. Y. (2020). Asian Americans report over 650 racist acts over last week, new data says. *NBC News*. https://www.nbcnews.com/news/asian-america/asian-americans-report-nearly-500-racist-acts-over-last-week-n1169821

Kapucu, N., Hu, Q., & Khosa, S. (2017). The state of network research in public administration. *Administration & Society*, *49*(8), 1087-1120.

Karni, A. (2020). In Daily Coronavirus Brief, Trump Tries to Redefine Himself. *New York Times*. https://www.nytimes.com/2020/03/23/us/politics/coronavirus-trump-briefing.html

Kettl, D. F. (2020). States Divided: The Implications of American Federalism for Covid-19. *Public Administration Review*.

Kogan, V., S. Lavertu, and Z. Peskowitz. (2016). Performance federalism and local democracy: Theory and evidence from school tax referenda. *American Journal of Political Science*, *60*(2), 418-435.

Lau, R. R. (1982). Negativity in political perception. *Political behavior*, *4*(4), 353-377.

Lodge, Milton, & Charles Taber (2000). "Three Steps Toward a Theory of Motivated Political
    33 Reasoning." In Arthur Lupia, Mathew McCubbins, & Samuel Popkin (Eds.), Elements
    of Reason: Cognition, Choice, and the Bounds of Rationality. New York: Cambridge
    University Press, pp. 183-213.

Maeder, E. M., Yamamoto, S., McManus, L. A., & Capaldi, C. A. (2016). Race– crime
    congruency in the Canadian context. *Canadian Journal of Behavioural Science/Revue
    canadienne des sciences du comportement*, *48*(2), 162.

Malhotra, N., and A.G. Kuo. (2008). Attributing blame: The public's response to Hurricane
    Katrina. *Journal of Politics*, *70*(1), 120–135.

Marvel, J. D. (2015b). Public opinion and public sector performance: Are individuals' beliefs
    about performance evidence-based or the product of anti–public sector bias?
    *International Public Management Journal*, *18*(2), 209-227.

Mettler, S. (2011). *The submerged state: How invisible government policies undermine
    American democracy*. University of Chicago Press.

Moynihan, D. P. (2008). *The Dynamics of Performance Management: Constructing Information
    and Reform*. Washington, DC: Georgetown University Press.

Moynihan, D. P., and J. Soss. (2014). Policy feedback and the politics of administration. *Public
    Administration Review*, *74*(3), 320-332.

Mummolo, J., & Peterson, E. (2019). Demand effects in survey experiments: An empirical
    assessment. *American Political Science Review*, *113*(2), 517-529.

Nabatchi, T., Sancino, A., & Sicilia, M. (2017). Varieties of participation in public services: The
    who, when, and what of coproduction. *Public Administration Review*, *77*(5), 766-776.

New York Times. (2020). The Illness Now Has a Name: COVID-19. *New York Times*.
    https://www.nytimes.com/2020/02/11/world/asia/coronavirus-china.html

Nielsen, P. A., and M. Baekgaard. (2015). Performance information, blame avoidance, and
    politicians' attitudes to spending and reform: Evidence from an experiment. *Journal of
    Public Administration Research and Theory*, *25*(2), 545–569.

Nielsen, P. A., and D. P. Moynihan. (2017). How do politicians attribute bureaucratic
    responsibility for performance? Negativity bias and interest group advocacy. *Journal of
    Public Administration Research and Theory*, *27*(2), 269-283.

Olsen, A. L. (2015). Citizen (dis)satisfaction: An experimental equivalence framing
    study. *Public Administration Review*, *75*(3), 469-478.

Petersen, M. B. (2012). Social welfare as small-scale help: evolutionary psychology and the deservingness heuristic. *American Journal of Political Science*, *56*(1), 1-16.

Petersen, N. B. G., Laumann, T. V., & Jakobsen, M. (2019). Acceptance or disapproval: Performance information in the eyes of public frontline employees. *Journal of Public Administration Research and Theory*, *29*(1), 101-117.

Pettigrew, T. F. (1979). The ultimate attribution error: Extending Allport's cognitive analysis of prejudice. *Personality and social psychology bulletin*, *5*(4), 461-476.

Redlawsk, D. P. (2002). Hot cognition or cool consideration? Testing the effects of motivated reasoning on political decision making. *The Journal of Politics*, *64*(4), 1021-1044.

Roberts, M. E., B. M. Stewart, D. Tingley, C. Lucas, J. LL., S. K. Gadarian, B. Albertson, and D.G. Rand. (2014). Structural topic models for open-ended survey responses. *American Journal of Political Science*, *58*(4), 1064-1082.

Rogers, K., L. Jakes, and A. Swanson. Trump Defends Using 'Chinese Virus' Label, Ignoring Growing Criticism. *New York Times*. https://www.nytimes.com/2020/03/18/us/politics/china-virus.html

Rozin, P., and Royzman, E. B.. (2001). Negativity bias, negativity dominance, and contagion. *Personality and Social Psychology Review*, *5*(4), 296–320.

Soroka, S. N. (2014). *Negativity in democratic politics: Causes and consequences*. Cambridge University Press.

Swim, J. K., & Sanna, L. J. (1996). He's skilled, she's lucky: A meta-analysis of observers' attributions for women's and men's successes and failures. *Personality and Social Psychology Bulletin*, *22*(5), 507-519.

Kelly, J. M., & Swindell, D. (2002). A multiple–indicator approach to municipal service evaluation: Correlating performance measurement and citizen satisfaction across jurisdictions. *Public administration review*, *62*(5), 610-621.

Taber, C. S., & Lodge, M. (2006). Motivated skepticism in the evaluation of political beliefs. *American journal of political science*, *50*(3), 755-769.

Tavernise, S. and R. Oppel Jr. (2020). Spit On, Yelled At, Attacked: Chinese-Americans Fear for Their Safety. *New York Times*. https://www.nytimes.com/2020/03/23/us/chinese-coronavirus-racist-attacks.html

Tilley, J., & Hobolt, S. B. (2011). Is the government to blame? An experimental test of how partisanship shapes perceptions of performance and responsibility. *The journal of politics*, *73*(2), 316-330.

Van den Bekerom, P., van der Voet, J., & Christensen, J. (2020). Are Citizens More Negative About Failing Service Delivery by Public Than Private Organizations? Evidence From a Large-Scale Survey Experiment. *Journal of Public Administration Research and Theory*.

Weaver, R. K. (1986). The politics of blame avoidance. *Journal of public policy*, 371-398.

Whitehead III, G. I., Smith, S. H., & Eichhorn, J. A. (1982). The effect of subject's race and other's race on judgments of causality for success and failure. *Journal of Personality*, *50*(2), 193-202

Zimmer, C. (2020). Most New York Coronavirus Cases Came from Europe Genomes Show. *New York Times*. https://www.nytimes.com/2020/04/08/science/new-york-coronavirus-cases-europe-genomes.html

**Appendix A: Vignettes**

**Baseline**

The Trump administration is dealing with the challenge of testing residents for a new and potentially dangerous virus. These days the performance of the Trump administration's performance is frequently discussed

**2x2 Treatment combines COVID-19 or Chinese virus with Positive and Negative Frame**

The Trump administration is dealing with the challenge of testing residents for a virus it has referred to as **COVID-19 [neutral description]/Chinese virus [scapegoat cue].** These days the Trump administration's performance is frequently discussed. Currently, the Trump administration is struggling to test people for COVID-19. The Trump administration estimates the United States **will [positive equivalence frame]/will not [negative equivalence frame]** have enough COVID-19 tests for **[random number between 51-99 for positive frame/1-49 for negative frame]** percent of people seeking tests in the next 10 days.

**Appendix B: Dependent Variables**

**Close-ended**

How would you evaluate the performance of the Trump administration in responding to this pandemic? (0 = Extremely Bad; 10 = Extremely Good)

How serious of a threat do you think the virus you just read about is? (0 = Completely overblown, 10 = Extremely Serious)

How responsible is each of the entities below for the spread of the virus you just read about in the United States? (0 = Not responsible at all; 10 = Extremely responsible)
Items: The Trump administration; the Obama administration; the Chinese government

How responsible is each of the entities below for the spread of the virus you just read about in the United States? (0 = Not responsible at all; 10 = Extremely responsible)
Items: Nancy Pelosi; State Governors; the mainstream media; the Centers for Disease Control

What ethnic group is most responsible for the spread of the virus you just read about in the United States
Select from: Chinese/Hispanic/Black/White/Korean

**Open-ended**

In the United States, who has done a bad job responding to the pandemic? Why?

**Motivated Reasoning and Blame: Responses to Performance Framing and Outgroup Triggers during COVID-19**

Supplementary Materials

The supplementary materials file consists of the following sections:

Section 1) Balance Test and Summary Statistics

Section 2) Results controlling for income

Section 3) Results for two-one sided t-tests (Equivalence testing)

Section 4) Structural topic model fit statistics

Section 5) Top 15 words for every topic in each topic model calculated using FREX

Section 6) Results using partisan affiliation (Republican) instead of partisan ideology (conservative)

*Section 1) Balance Test and Summary Statistics*

| | *Control* <br> *n = 287* | *Positively Framed Data* <br> *n = 566* | *Negatively Framed Data* <br> *n = 586* | **χ2/df,** <br> **p Value** | *Chinese Virus* <br> *n = 579* | *COVID-19* <br> *n = 573* | **χ2/df,** <br> **p Value** |
|---|---|---|---|---|---|---|---|
| *Gender%* | | | | | | | |
| Female | 46.34 | 48.23 | 46.49 | 1.603/4, $p = .808$ | 45.89 | 47.15 | .682/4, $p = .954$ |
| *Ethnicity%* | | | | | | | |
| American Indian | .36 | 0.512 | 0.707 | | 0.518 | 0.698 | |
| Asian | 6.97 | 9.556 | 9.894 | | 9.672 | 9.773 | |
| Black or African American | 9.76 | 12.116 | 12.544 | 6.43/12, $p = .893$ | 12.09 | 12.565 | 7.98/12, $p = .787$ |
| Hispanic | 4.87 | 4.437 | 4.417 | | 5.181 | 3.665 | |
| Native Hawaiian | .35 | 0.341 | 0.353 | | 0.173 | 0.524 | |
| Other | 1.74 | 2.048 | 1.06 | | 1.9 | 1.222 | |
| White | 75.95 | 70.99 | 71.025 | | 70.466 | 71.553 | |
| *Age (mean)* | | | | | | | |
| | 40.891 | 40.654 | 40.285 | 123.8/124, $p = .486$ | 40.057 | 40.880 | 106.46/124, $p = .870$ |
| *Income %* | | | | | | | |
| $10,000 - $19,999 | 7.666 | 7.244 | 5.802 | | 5.354 | 7.679 | |
| $100,000 - $149,999 | 10.453 | 11.484 | 11.433 | | 12.263 | 10.646 | |
| $20,000 - $29,999 | 13.24 | 10.424 | 8.532 | | 8.636 | 10.297 | |
| $30,000 - $39,999 | 11.847 | 12.191 | 10.58 | | 11.917 | 10.82 | |
| $40,000 - $49,999 | 8.014 | 11.661 | 11.775 | | 12.09 | 11.344 | |
| $50,000 - $59,999 | 9.756 | 10.777 | 12.628 | 34.66/22, $p = .04$ | 12.781 | 10.646 | 36.70/22, $p = 0.03$ |
| $60,000 - $69,999 | 11.15 | 7.597 | 7.679 | | 9.154 | 6.108 | |
| $70,000 - $79,999 | 8.014 | 6.007 | 10.41 | | 6.736 | 9.773 | |
| $80,000 - $89,999 | 4.53 | 6.537 | 4.949 | | 5.181 | 6.283 | |
| $90,000 - $99,999 | 4.53 | 6.36 | 7.85 | | 5.699 | 8.551 | |
| Less than $10,000 | 5.923 | 4.24 | 2.218 | | 3.454 | 2.967 | |
| More than $150,000 | 4.878 | 5.477 | 6.143 | | 6.736 | 4.887 | |
| *Political Affiliation %* | | | | | | | |
| Democrat | 44.948 | 40.813 | 46.075 | | 43.523 | 43.455 | |
| Independent | 32.056 | 28.622 | 24.573 | 10.72/6, $p = .09$ | 25.216 | 27.923 | 9.465/6, $p = .149$ |
| Other | 2.091 | 3.887 | 3.413 | | 4.318 | 2.967 | |
| Republican | 20.906 | 26.678 | 25.939 | | 26.943 | 25.654 | |
| *Education %* | | | | | | | |
| 2 year degree | 11.498 | 10.601 | 11.945 | | 11.917 | 10.646 | |
| 4 year degree | 43.902 | 46.113 | 44.881 | | 43.523 | 47.469 | |
| Graduate degree | 17.77 | 16.784 | 19.795 | 8.52/10, $p = .578$ | 18.48 | 18.15 | 4.05/10, $p = .945$ |
| High school graduate | 8.362 | 8.834 | 9.044 | | 9.154 | 8.726 | |
| Less than high school | 0.348 | 0.707 | 0 | | 0.518 | 0.175 | |

| | H1 | H2 | H3 | H4 | H5 | H6 |
|---|---|---|---|---|---|---|
| *Severity* | $F(1, 285) = 21.43, p < 0.000$ | - | - | - | - | - |
| *Perceptions of Trump's Performance* | $F(1, 285) = 90.52, p < 0.000$ | $F(1, 1128) = 0.08, p = 0.778$ | - | - | - | - |
| *Blame Obama* | $F(1, 285) = 11.64, p < 0.000$ | - | - | $F(1, 341) = 1.76, p = 0.196$ | - | $F(1, 1104) = 6.67, p = 0.01$ |
| *Blame Media* | $F(1, 285) = 9.34, p = 0.002$ | - | - | $F(1, 341) = 1.029, p = 0.311$ | - | - |
| *Blame Chinese Government* | $F(1, 285) = 14.39, p < 0.000$ | - | - | $F(1, 341) = 0.634, p = 0.426$ | $F(1, 1128) = 0.477, p = 0.489$ | $F(1, 1104) = 0.11, p = 0.74$ |
| *Blame Pelosi* | $F(1, 285) = 9.45, p = 0.002$ | - | - | $F(1, 341) = 1.521, p = 0.218$ | - | $F(1, 1104) = 5.23, p = 0.02$ |
| *Blame Trump* | $F(1, 285) = 51.29, p < 0.000$ | - | $F(1, 1128) = 0.07, p = 0.792$ | - | - | - |
| *Blame CDC* | $F(1, 285) = 2.82, p = 0.09$ | - | $F(1, 1150) = 1.01, p = 0.316$ | - | - | - |
| *Blame Chinese Residents* | $F(1, 285) = 12.81 p < 0.000$ | - | - | - | $F(1, 1128) = 6.29, p = 0.012$ | $F(1, 1148) = 0.81, p = 0.36$ |

H1: Motivated reasoning will lead to differences in how conservatives and non-conservatives assess the severity of the crisis, assess the performance of political leadership, and allocate responsibility to actors inside and outside of the United States Federal government.

H2: Exposure to negatively framed performance data will trigger more negative performance evaluations of actors in the United States federal government than exposure to positively framed performance data.

H3: Exposure to negatively framed performance data will trigger greater responsibility attribution to actors in the United States federal government than exposure to equivalent positively framed public health performance data.

H4: Conservatives exposed to the negatively framed performance information will be more likely to blame actors outside of the United States federal government when compared to those exposed to positively framed performance information.

H5: Participants exposed to the term 'Chinese virus' will be more inclined to blame Chinese residents and the Chinese government when compared to participants exposed to the term 'COVID - 19.'

H6: The effect of the "Chinese virus" framing on blaming Chinese residents and the Chinese government will be greatest for conservatives.

***Section 3) Results for two-one sided t-tests (Equivalence testing) for the Effect of Performance Data Framing***

      In this section we provide the results of equivalence tests for H2, H3, and H4 using two one-sided hypothesis tests using the TOSTER R package (Lakens 2017). In TOST equivalence testing, two independent t-tests are used to evaluate whether an effect differs significantly from equivalence bounds (or equivalence margin) that are equal to the smallest effect size of interest. Specifically, one t-test is used to examine whether an estimated effect differs significantly from the upper equivalence bound and a second to see whether the effect differs from the lower equivalence bound. If an effect falls within the upper and lower equivalence bounds (i.e., it is significantly smaller than the upper *and* significantly larger than the lower equivalence bound), we can accept the null hypothesis that an observed effect is too small to matter. If an effect falls within the upper or lower equivalence bounds (i.e., it is not significantly different from the upper *or* lower equivalence bound), we can reject the null, but not accept it (i.e., not conclude that the effect is too small to matter).

      For the figures below, we use equivalence bounds based on a small effect (Cohen's D = .2). A null hypothesis significance test at .05% corresponds to a 90% confidence interval for TOST, whereas a 95% confidence interval for TOST corresponds to a null hypothesis significance test at 0.025%.

**Reference**

Lakens, Daniël. 2017. Equivalence tests: A practical primer for t tests, correlations, and meta-analyses. *Social Psychological and Personality Science* **8**:355-362.

|  | *H2* | *H3* | *H4* |
|---|---|---|---|
| *Perceptions of Trump's Performance* | TOST Upper: $p = 0.045$<br>TOST Lower: $p = 0.001$<br>90% CI: -0.196, 0.472 | - | - |
| *Blame Obama* | - | - | TOST Upper: $p = 0.304$<br>TOST Lower: $p < 0.000$<br>90% CI: -0.116, 1.13 |
| *Blame Media* | - | - | TOST Upper: $p = 0.186$<br>TOST Lower: $p = 0.003$<br>90% CI: -0.261, 0.987 |
| *Blame Chinese Government* | - | - | TOST Upper: $p = 0.2157$<br>TOST Lower: $p = 0.002$<br>90% CI: -0.189, 0.875 |
| *Blame Pelosi* | - | - | TOST Upper: $p = 0.436$<br>TOST Lower: $p < 0.000$<br>90% CI: 0.017, 1.31 |
| *Blame Trump* | - | TOST Upper: $p < 0.000$<br>TOST Lower: $p < 0.000$<br>90% CI: -0.306, 0.334 | - |
| *Blame CDC* | - | TOST Upper: $p = 0.001$<br>TOST Lower: $p < 0.001$<br>90% CI: -0.116, 0.536 | - |

***Section 4) Structural topic model fit statistics***

As was explained in the main text, we selected a 5 topic topic-model, in part because of the relatively small corpus we use (Roberts, Stewart, and Tingley 2014). Additionally, a qualitative assessment of the words associated with each topic also indicated the basket of words for 5 topic topic-models was more coherent when compared to alternative 6 and 10 topic specifications. To add a more objective assessment of the number of topics to include in our structural topic models, we follow Bogozzi and Berliner (2016) and use model diagnostics as a piece of secondary, objective evidence in our evaluation. Specifically, to offer further rationale into the decision to use 5 topics, as opposed to, for example, 7 or 10 topic models, we quantitatively evaluate model fit diagnostics related to coherence and semantic cohesion. Semantic cohesion is an average that quantifies the degree to which words in each topic - across $n$ topics – relate to one another. Exclusivity reflects the degree to which the words in one topic are not present in other topics for a given structural topic model. In other words, this metric quantifies how distinct each topic is. Because semantic coherence and exclusivity vary in relation to the number of topics included in a structural topic model, the optimal number of topics reflects the number that scores highest in terms of semantic coherence and exclusivity. Below we report semantic coherence and exclusivity for 5, 7, and 10 topic models. As can be seen, in most cases the metrics favor a 5 topic structural topic model. In instances where the data suggest an alternative number of topics is optimal, the difference when compared to a 5 topic model is marginal. Results are provided below.

*Semantic Coherence and Exclusivity for Structural Topic model used for Hypothesis 1*

|  | *Semantic Coherence* | *Exclusivity* |
|---|---|---|
| *5 topic STM* | -67.68 | 6.99 |
| *7 topic STM* | -67.95 | 6.83 |
| *10 topic STM* | -86.87 | 6.33 |

*Semantic Coherence and Exclusivity for Structural Topic model used for Hypotheses 2 and 3*

|  | *Semantic Coherence* | *Exclusivity* |
|---|---|---|
| *5 topic STM* | -141.48 | 9.19 |
| *7 topic STM* | -157.34 | 9.36 |
| *10 topic STM* | -139.96 | 9.39 |

*Semantic Coherence and Exclusivity for Structural Topic model used for Hypotheses 4*

|  | *Semantic Coherence* | *Exclusivity* |
|---|---|---|
| *5 topic STM* | -119.17 | 7.18 |
| *7 topic STM* | -127.73 | 7.19 |
| *10 topic STM* | -126.84 | 6.61 |

*Semantic Coherence and Exclusivity for Structural Topic model used for Hypotheses 5*

|  | *Semantic Coherence* | *Exclusivity* |
|---|---|---|
| *5 topic STM* | -146.24 | 9.20 |
| *7 topic STM* | -150.14 | 9.35 |
| *10 topic STM* | -146.89 | 9.39 |

*Semantic Coherence and Exclusivity for Structural Topic model used for Hypotheses 6*

|  | *Semantic Coherence* | *Exclusivity* |
|---|---|---|
| *5 topic STM* | -117.50 | 7.18 |
| *7 topic STM* | -127.74 | 7.18 |
| *10 topic STM* | -127.47 | 6.62 |

*Section 5) Top 15 words for every topic in each topic model calculated using FREX*

| Hypothesis 1 | |
|---|---|
| *Topic Number* | *15 top words* |
| 1: President Response: General | peopl, presid, virus, feder, trump, mask, job, pandem, respond, wear, state, bad, donald, govern, governor |
| 2: Mask Confusion | job, donald, presid, pandem, mask, peopl, virus, state, wear, respond, trump, bad, feder, govern, governor |
| 3: States' Response | pandem, state, respond, governor, presid, job, peopl, trump, virus, mask, wear, bad, donald, feder, govern |
| 4: Mask Debate | govern, mask, wear, presid, virus, trump, job, peopl, pandem, state, respond, bad, donald, feder, governor |
| 5: President Response: Negative | trump, bad, presid, pandem, job, mask, virus, respond, peopl, wear, state, donald, feder, govern, governor |

| Hypotheses 2 and 3 | |
|---|---|
| *Topic Number* | *15 top words* |
| 1: President's Response | administr, peopl, donald, virus, spread, make, test, covid-19, time, terribl, slow, downplay, sever, follow, horribl |
| 2: Federal Government Response | job, bad, respond, pandem, feder, presid, poor, inform, media, situat, offici, local, horribl, act, terribl |
| 3:Democrats and Economic Downturn | economi, countri, distanc, trump, care, democrat, number, live, health, reopen, citizen, lie, governor, act, leader |
| 4: State Government | thing, social, govern, respons, good, made, polit, death, local, leadership, home, rule, american, blame, provid |
| 5: Mask Debate | mask, wear, state, public, governor, refus, open, case, unit, cdc, earli, action, busi, close, fail |

| Hypothesis 4 | |
|---|---|
| *Topic Number* | *15 top words* |
| 1: Administration | pandem, administr, respond, virus, bad, spread, mask, presid, peopl, democrat, govern, state, governor, job, media |
| 2: Liberal Left | bad, democrat, media, govern, spread, respond, state, peopl, mask, virus, pandem, presid, administr, governor, job |
| 3: Democrat's Response | job, state, govern, respond, spread, democrat, peopl, mask, virus, pandem, bad, presid, administr, governor, media |
| 4: Governors' Response | peopl, virus, governor, mask, state, govern, spread, democrat, bad, respond, pandem, presid, administr, job, media |
| 5: Partisanship | trump, presid, virus, state, democrat, pandem, peopl, respond, mask, spread, bad, govern, administr, governor, job |

| Hypothesis 5 | |
|---|---|
| *Topic Number* | *15 top words* |
| 1: President's Response | administr, donald, peopl, virus, spread, make, covid-19, time, test, downplay, slow, terribl, sever, follow, horribl |
| 2: Federal Gov's Response | job, bad, respond, feder, pandem, presid, media, poor, inform, offici, situat, act, local, horribl, test |
| 3: Democrats and Economic Downturn | economi, distanc, countri, care, democrat, trump, governor, number, live, health, reopen, lie, citizen, leader, ignor |
| 4: Government in General | social, thing, made, govern, good, respons, polit, local, home, death, leadership, rule, american, blame, provid |
| 5: Mask Debate | mask, wear, state, refus, public, open, unit, case, cdc, action, earli, busi, close, fail, governor |

| Hypothesis 6 | |
|---|---|
| *Topic Number* | *15 top words* |
| 1: President's Response | president, administration, virus, people, pandemic, states, government, media, bad, job, state, trump |
| 2: Everyone but the President | people, bad, media, pandemic, states, virus, government, president, administration, job, state, trump |
| 3: States' Response | job, pandemic, states, people, media, virus, government, president, administration, bad, state, trump |
| 4: Trump's Response | trump, virus, people, states, pandemic, media, government, president, administration, bad, job, state |
| 5: Government in General | government, state, people, media, states, president, pandemic, virus, administration, bad, job, trump |

| | H1 | H2 | H3 | H4 | H5 | H6 |
|---|---|---|---|---|---|---|
| *Severity* | $F(1, 285) = 7.88, p < 0.005$ | - | - | - | - | - |
| *Perceptions of Trump's Performance* | $F(1, 285) = 86.10, p < 0.000$ | $F(1, 1148) = 0.48, p = 0.496$ | | - | - | - |
| *Blame Obama* | $F(1, 285) = 24.5, p < 0.000$ | | - | $F(1, 301) = 1.29, p = 0.257$ | - | $F(1, 301) = 0.46, p = 0.498$ |
| *Blame Media* | $F(1, 285) = 16.3, p < 0.001$ | | - | $F(1, 301) = 1.30, p = 0.255$ | - | |
| *Blame Chinese Government* | $F(1, 285) = 27.5, p < 0.000$ | - | | $F(1, 301) = 0.379, p = 0.539$ | $F(1, 1148) = 0.334, p = 0.563$ | $F(1, 301) = 0.289, p = 0.59$ |
| *Blame Pelosi* | $F(1, 285) = 19.00, p <0.000$ | - | - | $F(1, 301) = 4.70, p = 0.03$ | - | $F(1, 301) = 0.608, p = 0.436$ |
| *Blame Trump* | $F(1, 285) = 30.10, p < 0.000$ | - | $F(1, 1148) = 0.004, p = 0.948$ | - | - | - |
| *Blame CDC* | $F(1, 285) = 7.78, p = 0.005$ | - | $F(1, 1148) = 1.11, p = 0.291$ | - | - | - |
| *Blame Chinese Residents* | $F(1, 285) = 27.3, p < 0.000$ | - | - | - | $F(1, 1148) = 6.40, p = 0.012$ | $F(1, 301) = 0.205, p = 0.65$ |

H1: Motivated reasoning will lead to differences in how Republicans and non Republicans assess the severity of the crisis, assess the performance of political leadership, and allocate responsibility to actors inside and outside of the United States Federal government.

H2: Exposure to negatively framed performance data will trigger more negative performance evaluations of actors in the United States federal government than exposure to positively framed performance data.

H3: Exposure to negatively framed performance data will trigger greater responsibility attribution to actors in the United States federal government than exposure to equivalent positively framed public health performance data.

H4: Republicans exposed to the negatively framed performance information will be more likely to blame actors outside of the United States federal government when compared to those exposed to positively framed performance information.

H5: Participants exposed to the term 'Chinese virus' will be more inclined to blame Chinese residents and the Chinese government when compared to participants exposed to the term 'COVID - 19.'

H6: The effect of the "Chinese virus" framing on blaming Chinese residents and the Chinese government will be greatest for Republicans.

***Results using partisan affiliation (Republican) instead of partisan ideology (conservative) for open-ended items.***

*H1: Motivated reasoning will lead to differences in how Republicans and non Republicans assess the severity of the crisis, assess the performance of political leadership, and allocate responsibility to actors inside and outside of the United States Federal government.*

For hypothesis 1, topics, topic prevalence, and variation attributable to partisan affiliation does not vary substantively from the results in the main text.

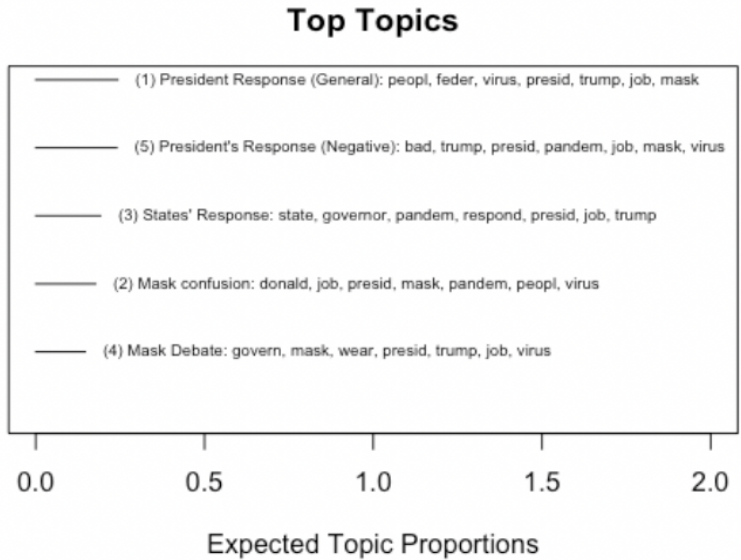

Figure 1a: 5 Most frequent topic proportions among all participants in the control group for the question *Who do you think did a bad job responding to the pandemic?*

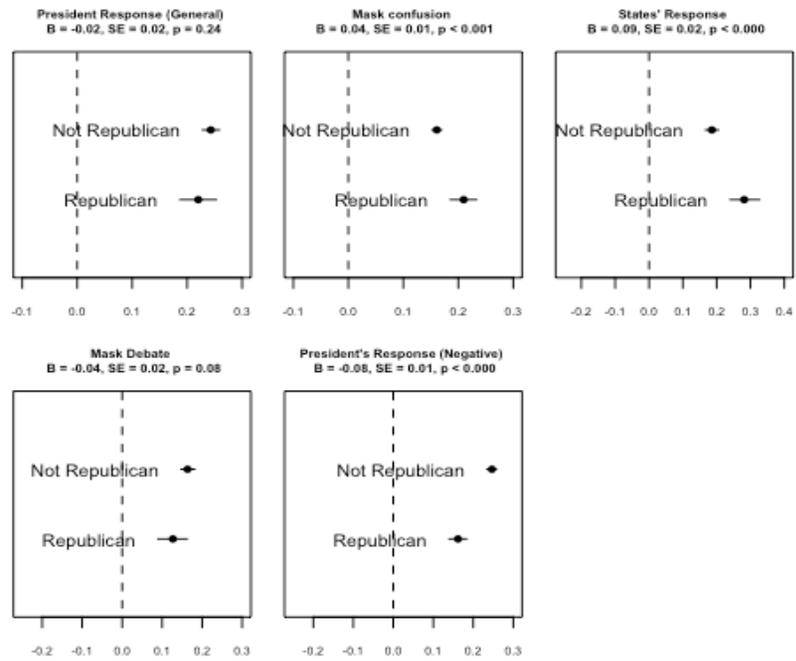

Figure 1b: Graphical display of topical prevalence contrast between Republicans and non-Republicans in the control group for the question *Who do you think did a bad job responding to the pandemic?* Horizontal lines represent 95% confidence intervals.

*H4: Exposure to negatively framed performance data will trigger more negative performance evaluations of actors in the United States federal government than exposure to positively framed performance data.*

While the specific names of the topics differ from those in the main text, as well their prevalence results in 2b shows the pattern of effects that results from exposure to performance information framing on blame attribution correspond to the results reported in the main text.

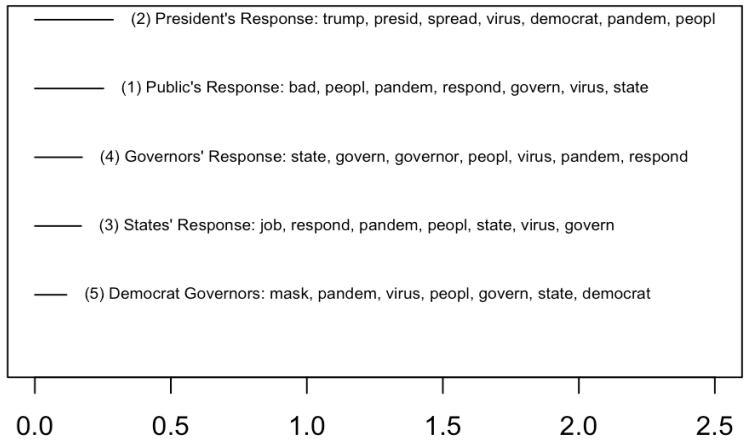

Figure 2a: Most frequent topic proportions among Republicans for the question *Who do you think did a bad job responding to the pandemic?* In estimating these topics, prevalence is estimated using exposure to performance information framing.

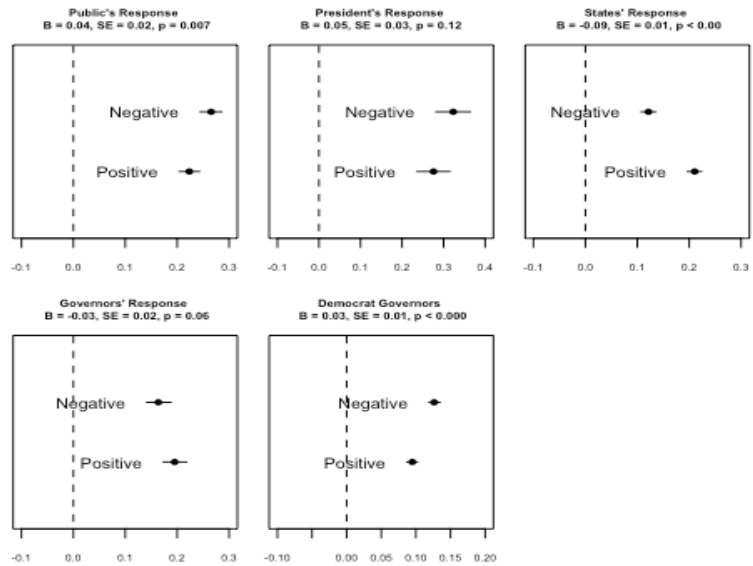

Figure 2b: Graphical display of topical prevalence contrast between exposure to positively and negatively framed performance information among Republicans for responses to the question *Who do you think did a bad job responding to the pandemic?* Horizontal lines represent 95% confidence intervals.

*H6: The effect of the "Chinese virus" framing on blaming Chinese residents and the Chinese government will be greatest for Republicans.*

While the specific names of the topics differ from those in the main text, as well their prevalence results in 2b shows the impact of performance information framing on blame attribution correspond to the results reported in the main text. Namely, we observe that use of the term 'Chinese virus' does result in Republican participants blaming the public, it also shifts blame away from actors President Trump frequently criticized: Democrat governors, governors in general, and states governments.

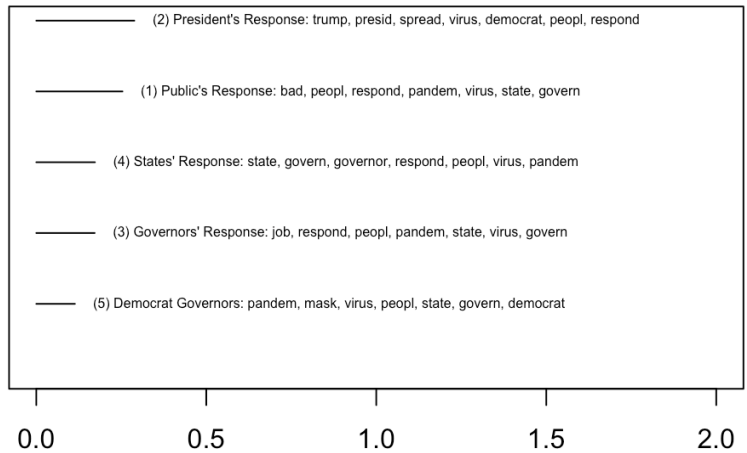

Figure 3a: 5 Most frequent topic proportions among Republican participants assigned to the Chinese virus and COVID-19 treatment for the question *Who do you think did a bad job responding to the pandemic?* In estimating these topics, prevalence is estimated using the scapegoat treatment.

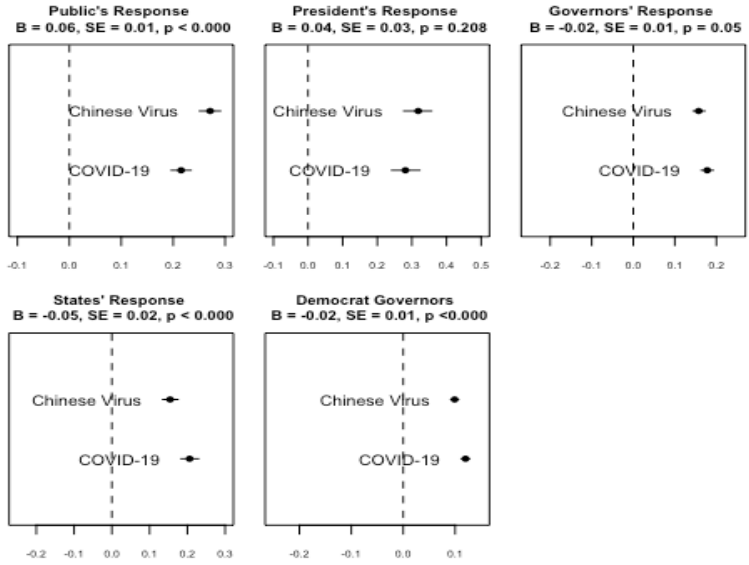

Figure 3b: Graphical display of topical prevalence contrast between Republicans exposed to the term Chinese virus and COVID-19 for the question *Who do you think did a bad job responding to the pandemic?* Horizontal lines represent 95% confidence intervals.